\documentclass[preprintnumbers,floatfix,letterpaper,aps,prd,nofootinbib,onecolumn]{revtex4-2}
\usepackage{amsmath,amsfonts,latexsym,amssymb,graphicx,graphics,epsfig,subfigure,color,makeidx,array}
\usepackage{xcolor,diagbox}
\usepackage{multirow}
\usepackage[colorlinks,linkcolor=blue,anchorcolor=blue,citecolor=blue,urlcolor=blue]{hyperref}
\usepackage{mathrsfs}
\usepackage{amsmath}
\usepackage{bm,graphicx,dcolumn,epstopdf,epsf,latexsym,mathbbol, amssymb,amsmath,color,slashed, mathrsfs,mathcomp,simplewick}
\usepackage{makecell}

\newcommand{\bq}{\begin{equation}}
\newcommand{\eq}{\end{equation}}
\newcommand{\bqn}{\begin{eqnarray}}
\newcommand{\eqn}{\end{eqnarray}}

\newcommand{\lb}{\label}
\newcommand{\f}{\frac} 
\newcommand{\p}{\partial}
\newcommand{\tx}{\text}

\newcommand{\lf}{\left}
\newcommand{\rt}{\right}

\begin{document}

\title{Constraining polymerized black holes with quasi-circular extreme mass-ratio inspirals}

\author{Sen Yang${}^{a, b}$}
\email{120220908881@lzu.edu.cn}

\author{Yu-Peng Zhang${}^{a, b}$}
\email{zhangyupeng@lzu.edu.cn}

\author{Tao Zhu${}^{c, d}$}
\email{zhut05@zjut.edu.cn}

\author{Li Zhao${}^{a, b}$}
\email{lizhao@lzu.edu.cn, corresponding author}

\author{Yu-Xiao Liu${}^{a, b}$}
\email{liuyx@lzu.edu.cn, corresponding author}

\affiliation{
${}^{a}$ Lanzhou Center for Theoretical Physics, \\ 
Key Laboratory of Theoretical Physics of Gansu Province,\\
Key Laboratory of Quantum Theory and Applications of the Ministry of Education,\\
Gansu Provincial Research Center for Basic Disciplines of Quantum Physics,
Lanzhou University, Lanzhou 730000, China\\
${}^{b}$ School of Physical Science and Technology,\\
Institute of Theoretical Physics $\&$ Research Center of Gravitation, Lanzhou University, Lanzhou 730000, China\\
${}^{c}$ Institute for Theoretical Physics and Cosmology, Zhejiang University of Technology, Hangzhou, 310023, China\\
${}^{d}$ United Center for Gravitational Wave Physics (UCGWP), Zhejiang University of Technology, Hangzhou, 310023, China\\}

\begin{abstract}
In this paper, we focus on the gravitational waves emitted by a stellar-mass object in a quasi-circular inspiral orbit around a central supermassive polymerized black hole in loop quantum gravity. Treating the stellar-mass object as a massive test particle, we derive its equations of motion and the corresponding radial effective potential. We find that the peak of the radial effective potential decreases with the quantum parameter $\hat{k}$. We also examine the impact of quantum corrections on the properties of stable circular orbits around the polymerized black hole. \textcolor{black}{We model the smaller object's trajectory as an adiabatic evolution along stable circular orbits using a semi-relativistic approach. In this method, the motion of the object is described by relativistic geodesics, and changes in the object's energy and orbital angular momentum due to gravitational radiation are calculated using the mass quadrupole formula.}~The corresponding gravitational waveforms are generated using the numerical kludge method, revealing that quantum corrections cause phase advances in the gravitational waveforms. We further analyze the potential constraints on the quantum parameter $\hat{k}$ from future space-based gravitational wave observations, concluding that these observations will likely impose stronger constraints on $\hat{k}$ than those obtained from black hole shadow measurements.

\end{abstract}


\maketitle

\section{Introduction}
\label{Introduction}
\renewcommand{\theequation}{1.\arabic{equation}} 
\setcounter{equation}{0}

The Event Horizon Telescope made a significant breakthrough by capturing the first image of the supermassive black hole in the M87 galaxy in 2019 \cite{EventHorizonTelescope:2019dse}, followed by the first image of the supermassive black hole at the center of the Milky Way announced in 2022 \cite{EventHorizonTelescope:2022wkp}. These images provide direct visual evidence of supermassive objects at the centers of galaxies. In each galaxy, there are many stellar-mass objects orbiting around the central supermassive black hole. Over time, these stellar-mass objects slowly inspiral inward the central supermassive black hole due to the gravitational radiation, which are extreme mass-ratio inspiral (EMRI) systems \cite{Hughes:2000ssa}. The gravitational waves emitted by these EMRIs are expected to be detected by future space-based gravitational wave detectors, such as Laser Interferometer Space Antenna (LISA) \cite{LISA:2017pwj}, Taiji \cite{Hu:2017mde}, TianQin \cite{TianQin:2015yph}, and DECi-hertz
Gravitational-wave Observatory (DECIGO) \cite{Musha:2017usi}. 

For each EMRI, the smaller object undergoes numerous orbits around the central supermassive black hole before ultimately plunging into it \cite{Hughes:2000ssa}. The orbital dynamics of the smaller object are mainly determined by the properties of the central supermassive black hole and gravitational radiation. The gravitational waveforms emitted by an EMRI are closely related to the trajectory of the smaller object, serving as precise mappings of the spacetime geometry around the supermassive black hole. Thus, gravitational waves from EMRIs are invaluable probes for studying the characteristics of supermassive black holes and testing gravitational theories \cite{Glampedakis:2005hs, Barausse:2020rsu, LISA:2022yao, GWSTQLISA, Babak:2017tow,Qiao:2024gfb}. 

\textcolor{black}{The semi-relativistic
methods (such as analytical and numerical kludge methods~\cite{Barack:2003fp, Babak:2006uv, Sopuerta:2011te, Chua:2017ujo, Liu:2020ghq}) are effective for studying the evolutions of EMRIs. The numerical kludge (NK) method combines the exact relativistic orbital trajectory of the smaller object with an approximate expression for the gravitational radiation \cite{Babak:2006uv}.} Typically, the orbit of the smaller object is elliptical. However, as gravitational radiation leads to a gradual loss of energy and orbital angular momentum, the eccentricity of the orbit gradually decreases~\cite{Peters:1964zz}. Gravitational radiation from a particle in a circular orbit around a black hole was studied in Refs.~\cite{Poisson:1993vp, Cutler:1993vq, Apostolatos:1993nu}. To describe the orbital evolution of the smaller object, the quasi-circular orbit approximation is often employed \cite{Kennefick:1995za}. In this approximation, the smaller object is assumed to remain in a circular orbit at each moment, with the radius of the orbit slowly decreasing over time due to gravitational radiation. References~\cite{Kennefick:1995za, Finn:2000sy, Hughes:1999bq, Hughes:2001jr} employed the quasi-circular orbit approximation to study the gravitational waves from a compact object orbiting a supermassive Kerr black hole. The quasi-circular orbit approximation was also used to study LISA's capability
to detect new fundamental fields in Refs.~\cite{Maselli:2021men, Barsanti:2022vvl,Zhang:2024ogc}. More recently, quasi-circular EMRIs have been used to investigate the properties of black holes in bumblebee gravity in Ref. \cite{Liang:2022gdk}.

The existence of black holes is a key prediction of general relativity. However, the singularities predicted within black holes pose a significant theoretical challenge. As a potential solution, loop quantum gravity has proposed several black hole models that eliminate these singularities \cite{Rovelli:1997yv, Ashtekar:2021kfp, Perez:2017cmj}. References~\cite{Tu:2023xab, Yang:2024lmj, Liu:2024qci, Fu:2024cfk} have explored the EMRIs around several black holes in loop quantum gravity. However, Refs.~\cite{Tu:2023xab, Yang:2024lmj} examined EMRIs over only a few orbital cycles, neglecting the effects of gravitational radiation on the orbital evolution of the smaller object. Through the semi-classical polymerization of the area within the interior of spherically symmetric black hole spacetimes, a quantum-corrected black hole spacetime has been proposed in Refs.~\cite{Peltola:2008pa, Peltola:2009jm}. Various properties of this polymerized black hole have been studied, including Hawking radiation \cite{Fen-Fen:2013xga}, quasinormal modes \cite{Daghigh:2020mog, Jha:2023rem}, strong gravitational lensing and shadow \cite{Jha:2023rem, KumarWalia:2022ddq}, and so on.

In this work, we focus on gravitational waveforms generated by a stellar-mass object inspiraling into a supermassive polymerized black hole along the quasi-circular orbits. Treating the smaller object as a massive test particle, we derive its equations of motion and obtain the corresponding radial effective potential. We investigate the impact of the quantum corrections on stable circular orbits (SCOs), including the innermost stable circular orbits (ISCOs), around the polymerized black hole.~Then, we employ a numerical algorithm to model the quasi-circular orbit inspirals. We explore how the quantum corrections affect the evolution of the quasi-circular inspirals. Considering a stellar-mass object moving along the quasi-circular orbits around a supermassive polymerized black hole, we analyze the gravitational waveforms from this EMRI system by using the numerical kludge scheme \cite{Babak:2006uv} and discuss how the quantum corrections influence these waveforms. We further account for the Doppler shift induced by LISA's motion~\cite{Barack:2003fp} and calculate the resulting gravitational wave response signal in LISA. To assess the observability of the quantum parameter's effects on gravitational waves, we calculate the mismatch between waveforms for different values of the quantum parameter and those of a Schwarzschild black hole. Finally, we investigate the critical value of the quantum parameter, below which the gravitational waves from quasi-circular EMRIs of the polymerized black hole become indistinguishable from those of a Schwarzschild black hole.

This paper is organized as follows. In Sec.~\ref{sec2}, we analyze the geodesic of a massive test particle on the equatorial orbit around the polymerized black hole, derive the radial effective potential, and investigate the properties of the general SCOs and the ISCOs.  In Sec.~\ref{sec3}, we adopt a numerical algorithm to model the quasi-circular orbit inspirals. Then we study the gravitational waveforms of the test object along quasi-circular orbits around the supermassive polymerized black hole, investigate their dephasing, and explore their distinguishability in Sec.~\ref{sec4}. Finally, the conclusions and discussions of this work are given in Sec.~\ref{sec5}. Throughout the paper, we use the geometrized unit system with $G = c = 1$.

\section{Timelike geodesics} 
\label{sec2}
\renewcommand{\theequation}{2.\arabic{equation}} 
\setcounter{equation}{0}

Applying the polymerization of the area operator in the interior of a black hole, a new black hole metric in LQG has been proposed in Refs.~\cite{Peltola:2008pa, Peltola:2009jm}. This polymerized black hole offers potential insights into the quantum dynamics inside black holes and could help in understanding the connection between quantum gravity and general relativity. The line element of the polymerized black hole is  
\bqn\lb{metric}
ds^2 = - \left(\sqrt{1- \frac{k^2}{r^2}} - \frac{2M}{r} \right) dt^2 + \frac{dr^2}{\left( \sqrt{1- \frac{k^2}{r^2}} - \frac{2M}{r} \right) \left( 1- \frac{k^2}{r^2} \right)} + r^2 d \theta^2 + r^2 \sin ^2 \theta d \varphi ^2,
\eqn 
where $M$ is the Arnowitt–Deser–Misner mass and $k$ is the quantum-corrected parameter which determines the minimal radius of the spacetime.  The radius of the event horizon for the polymerized black hole is $\sqrt{4 M^2 + k^2}$. Equation \eqref{metric} returns to the line element of a Schwarzschild black hole when $k = 0$. \textcolor{black}{For convenience, we use the dimensionless parameter $ \hat{k} = k/M$ instead of $k$ in this work. Through the observations of M~87* and Sgr A* black hole shadow from the Event Horizon Telescope, Ref.~\cite{KumarWalia:2022ddq} obtained constraints on $k$ and the strongest one is $\hat{k} \leq 0.36$.}

\textcolor{black}{It is important to note that the true value of the parameter $\hat{k}$ is expected to be extremely small. In an EMRI system, a smaller object orbits a central supermassive black hole over many years, and very long-duration gravitational wave signals are generated. Gravitational waves from EMRIs can map the background spacetime geometry with high precision. Although the value of the parameter $\hat{k}$ is small, its effects may still become detectable over extended observation periods. If future space-based gravitational wave detectors detect gravitational waves from EMRIs, we could use these gravitational waves to derive precise constraints on the black hole metric parameters. In this work we use a significantly larger value for $\hat{k}$, so that we can better highlight its impact on the gravitational wave signal and ensure that its influence significantly exceeds numerical errors.}

\begin{figure}[!t]
	\centering
	\subfigure[]{\includegraphics[scale =0.28]{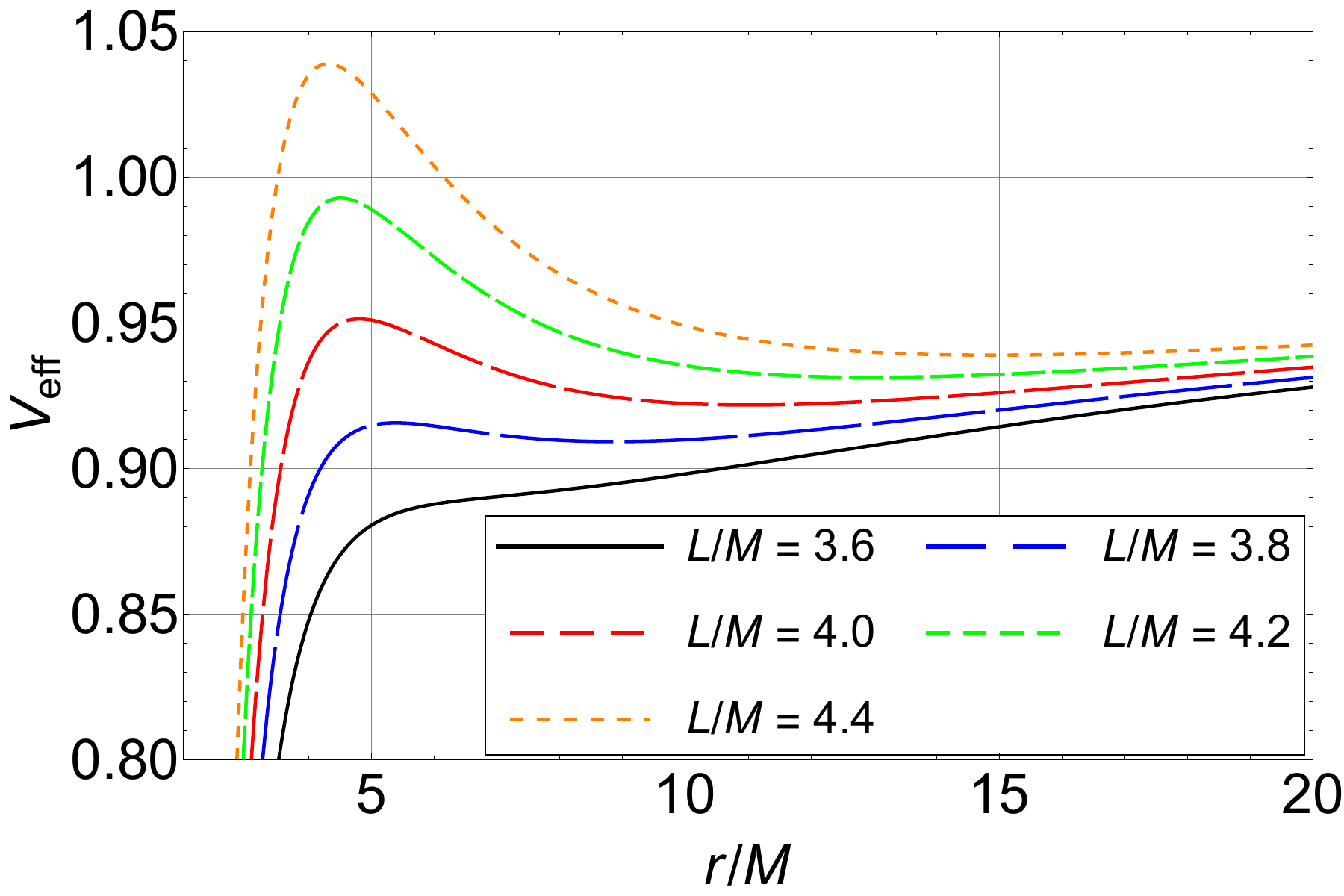} \lb{plot-V-1}}
	\subfigure[]{\includegraphics[scale =0.28]{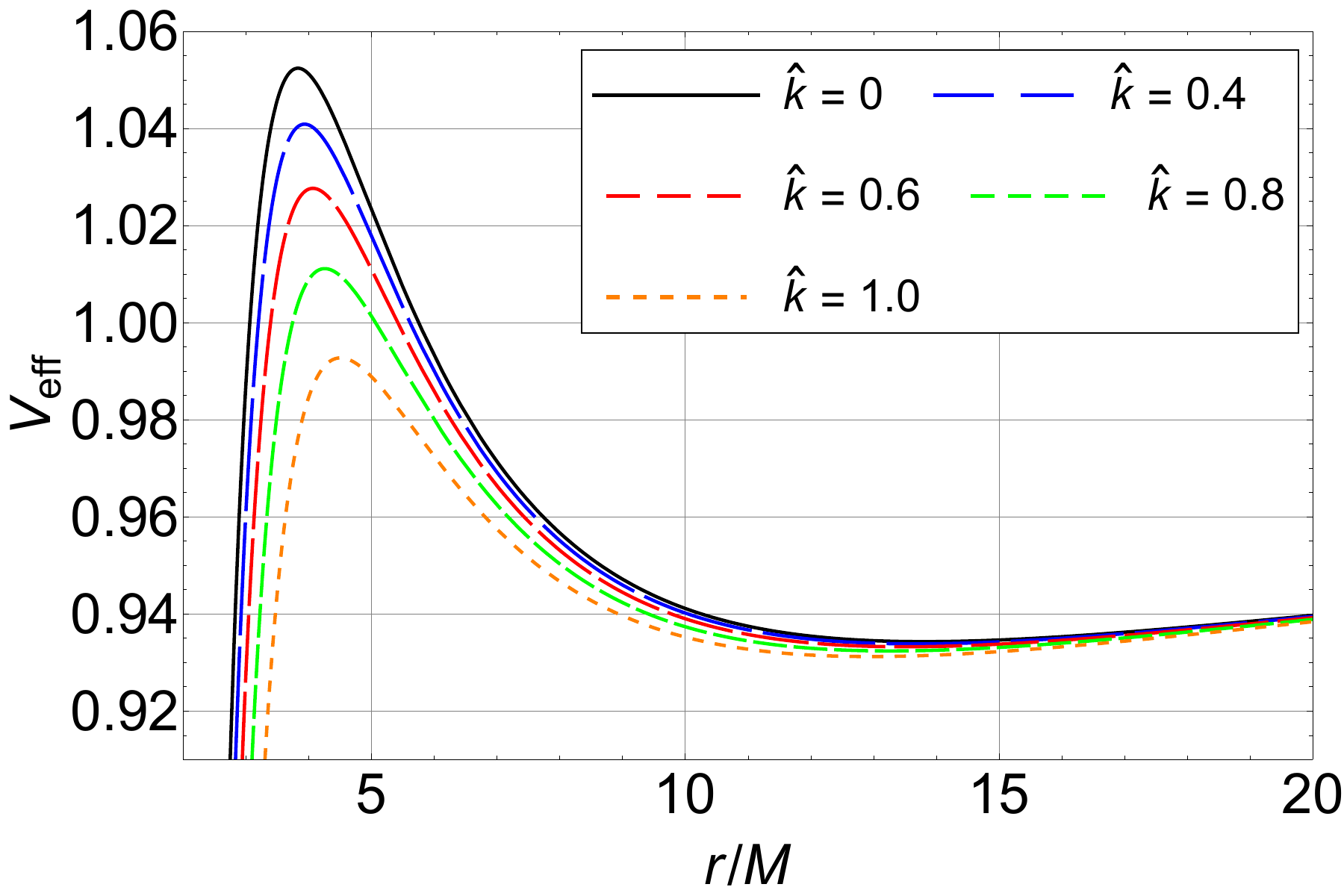} \lb{plot-V-2}}
	\caption{(a) The radial effective potentials of the test particle with different values of the orbital angular momentum around the polymerized black hole with $\hat{k} =1$. (b) The radial effective potentials of the test particle with fixed orbital angular momentum $L/M = 4.2$ around the polymerized black hole with different values of the parameter $\hat{k}$. }
	\label{plot-V}
\end{figure}

To explore the properties of the polymerized black hole, we study an EMRI system consisting of a stellar-mass object orbiting a central supermassive polymerized black hole. The gravitational waves from this EMRI system will provide a highly accurate probe of the polymerized black hole's spacetime geometry. The gravitational waveforms from the EMRI system are closely related to the trajectory of the smaller object. When the influence of the smaller object on the spacetimes is ignored, we can treat the smaller object as a massive test particle and study its orbit with geodesic at every moment. \textcolor{black}{We use the Lagrangian method \cite{Chandrasekhar:1985kt} to derive the equations of motion for the test particle moving on the equatorial plane $(\theta = \pi/2)$ around the polymerized black hole in Appendix \ref{AppA}.} 
Eqution~\eqref{Lagrangian-2} can be rewritten as
\bqn\lb{Veff-E}
\f{r^2}{r^2 - k^2} \dot{r}^2 + V_{\tx{eff}} = E^2,
\eqn
where the radial effective potential is 
\bqn\lb{Veff}
V_{\tx{eff}} = \lf(\sqrt{1- \frac{k^2}{r^2}} - \frac{2M}{r} \rt) \lf( 1+\f{L^2}{r^2} \rt). 
\eqn
We plot the radial effective potential \eqref{Veff} in Fig.~\ref{plot-V} to investigate its properties. As shown in Fig~\ref{plot-V-1}, the radial effective potential will possess extremum with the increase of the orbital angular momentum. The minimum of the radial effective potential corresponds to the stable circular orbits and the maximum corresponds to the unstable ones. Figure \ref{plot-V-2} shows the maximum of the radial effective potential decreases with the parameter $\hat{k}$. 

\begin{figure}[!t]
	\centering
	\subfigure[]{\includegraphics[scale =0.28]{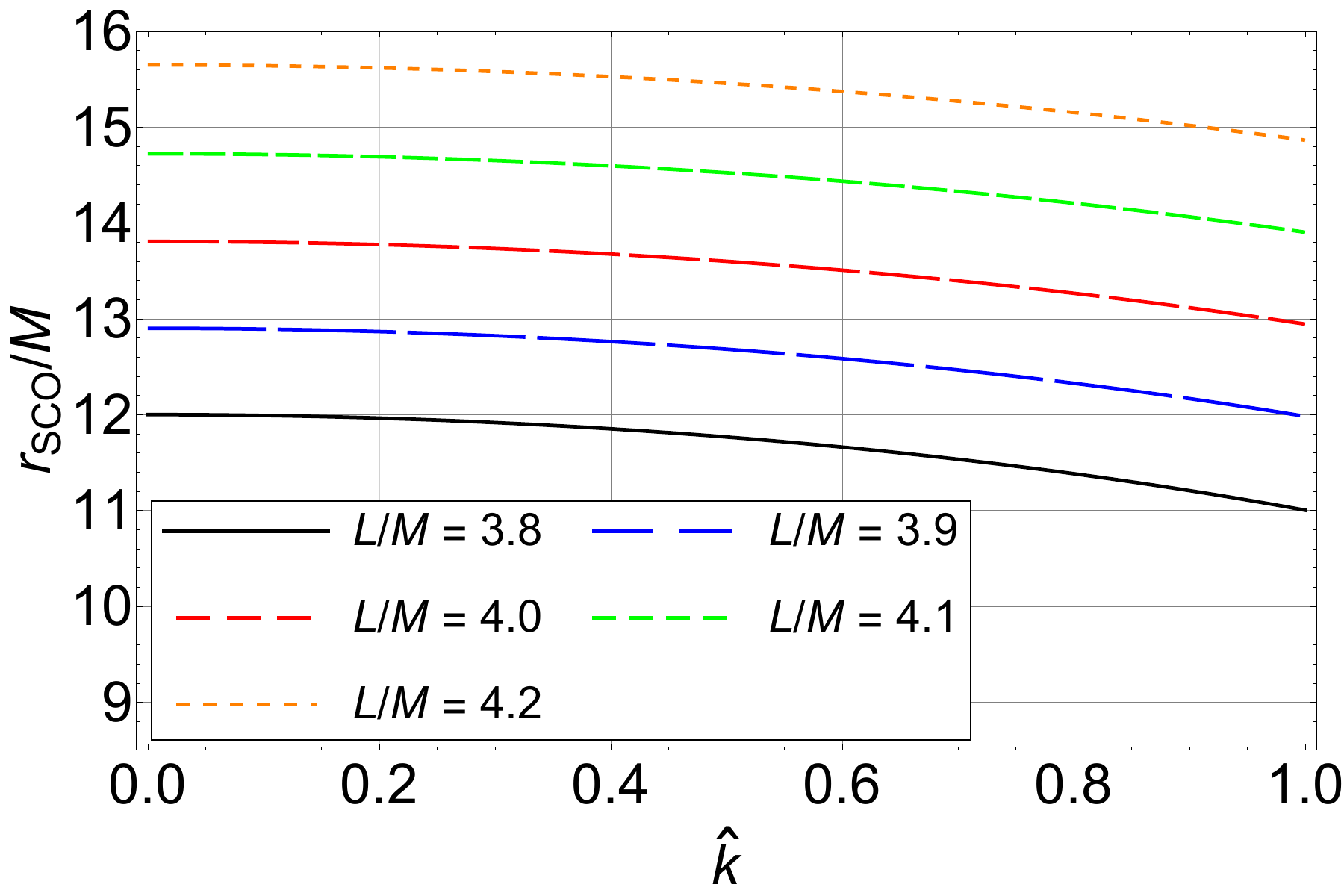}}
	\subfigure[]{\includegraphics[scale =0.28]{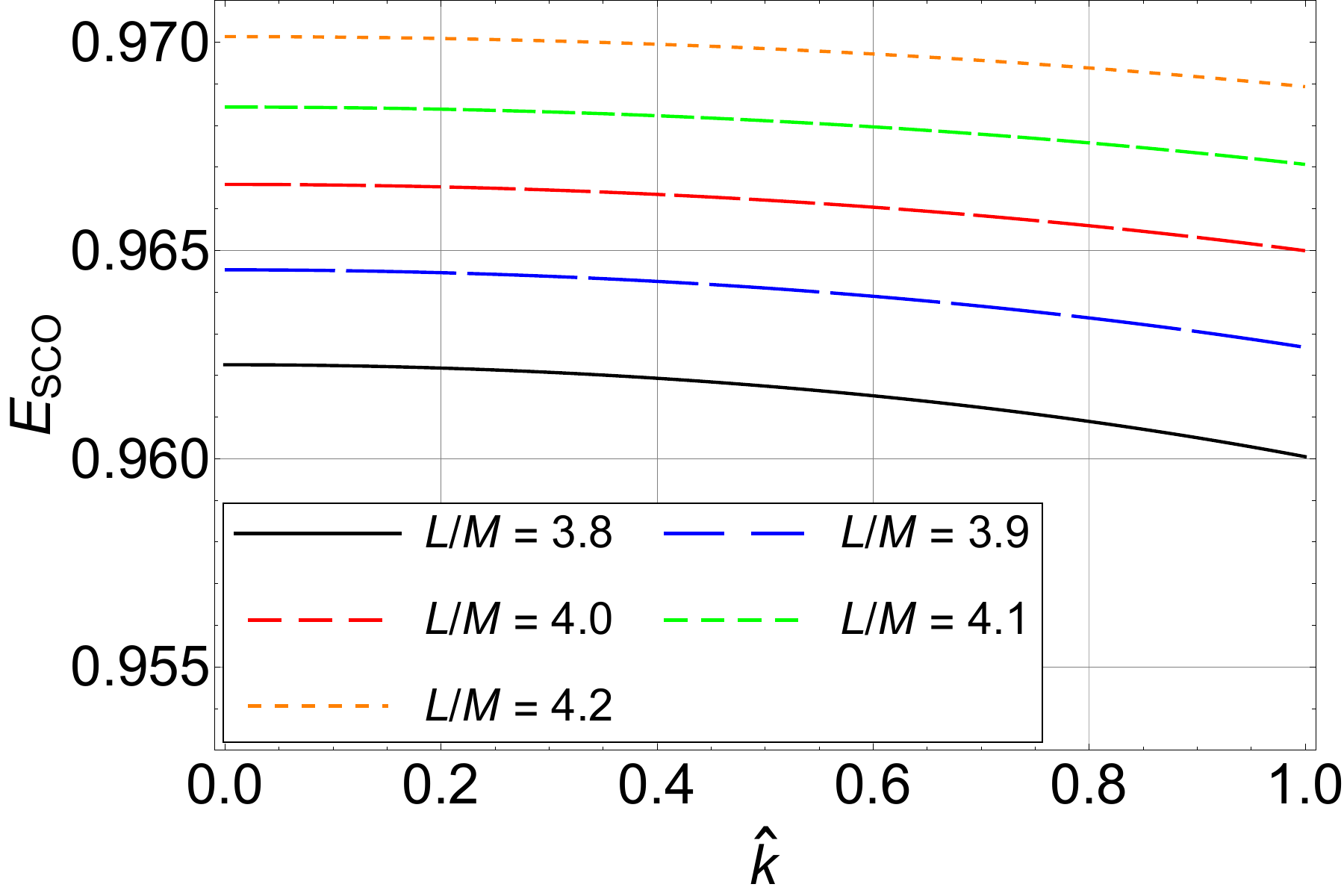}}
	\caption{(a) The radius of the SCO around the polymerized black hole as a function of the parameter $\hat{k}$. (b) The energy of the test particle along SCO around the polymerized black hole as a function of the parameter $\hat{k}$.}
	\label{plot-SCO}
\end{figure}

\begin{figure}[!t]
	\centering
	\subfigure[]{\includegraphics[scale =0.25]{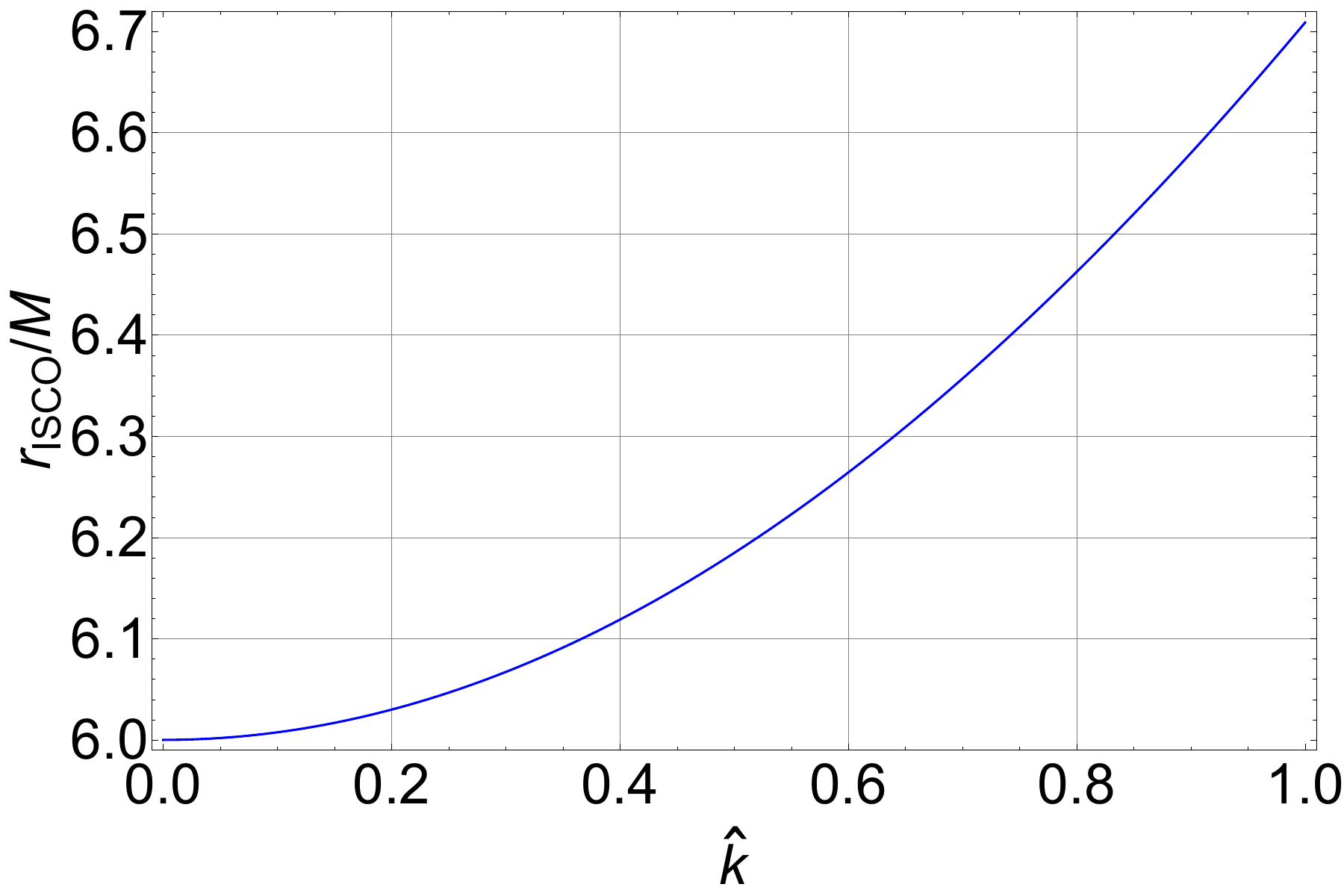}}
	\subfigure[]{\includegraphics[scale =0.25]{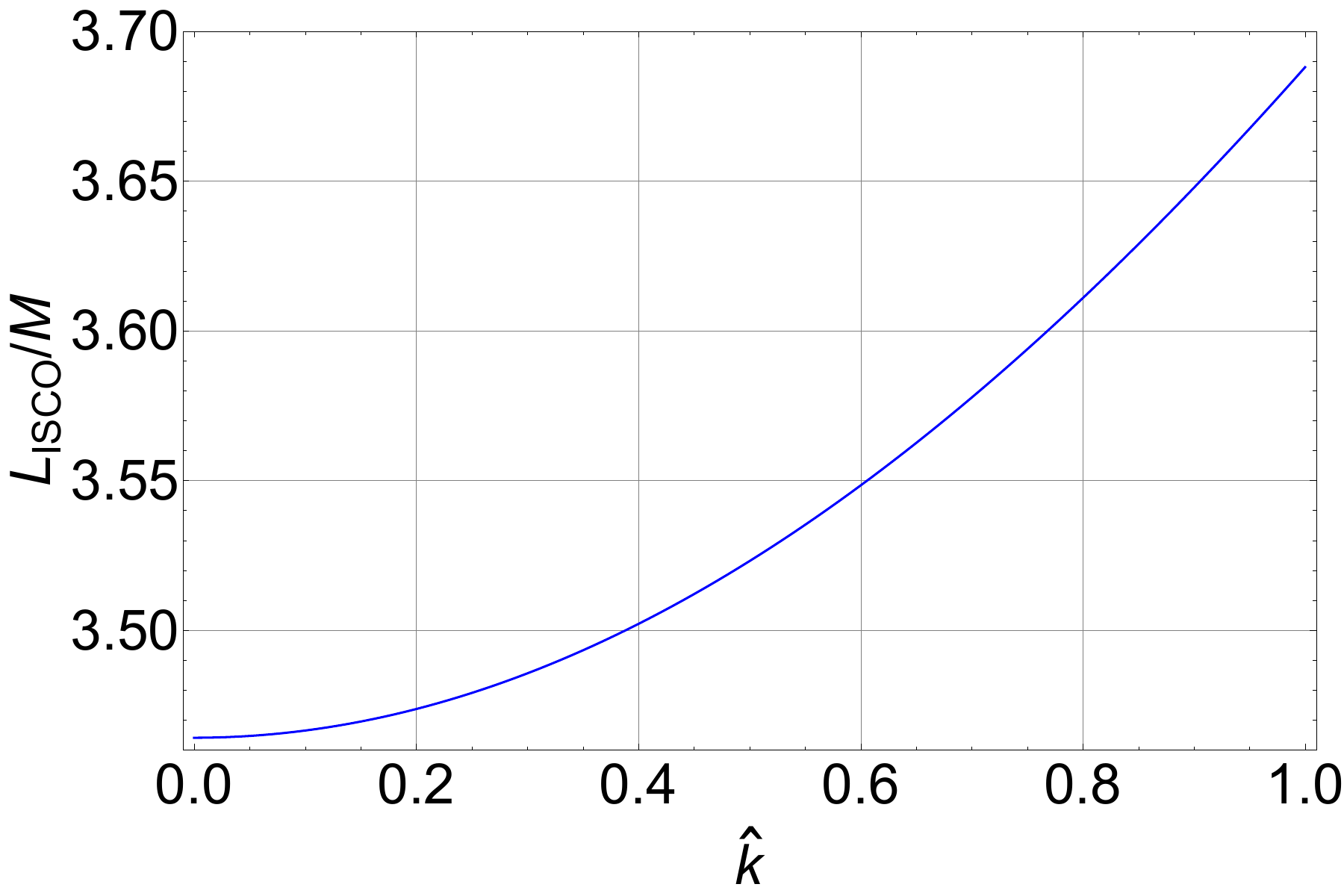}}
    \subfigure[]{\includegraphics[scale =0.25]{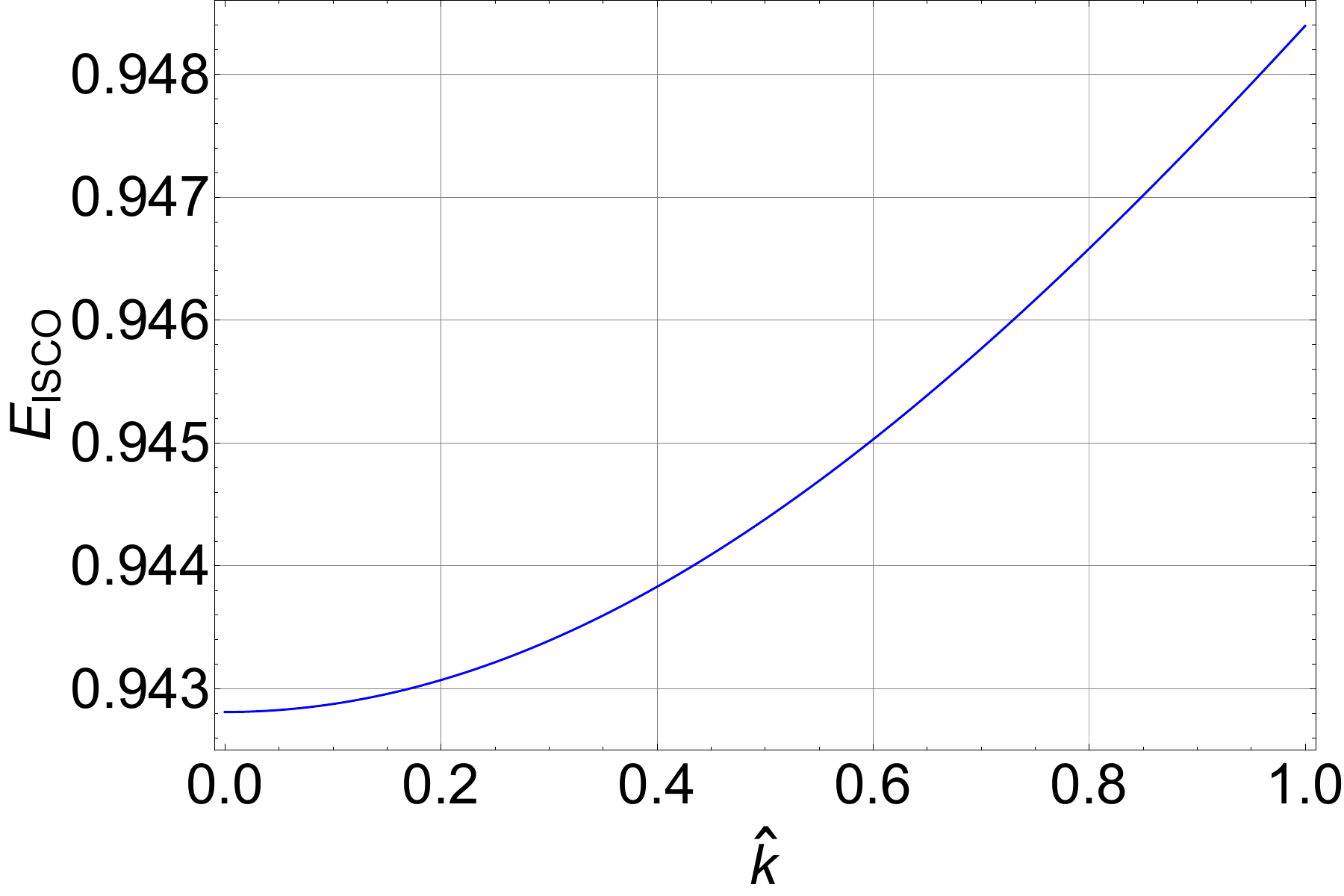}}
	\caption{The radius (a), the orbital angular momentum (b), and the energy (c) of the ISCO around the polymerized black hole as functions of the parameter $\hat{k}$.}
	\label{plot-ISCO}
\end{figure}

In the late stage of an EMRI, the eccentricity of the smaller object's orbit is extremely small, nearly zero~\cite{Peters:1964zz}. So the assumption of a quasi-circular orbit is a reasonable approximation in modeling EMRIs' orbital evolution~\cite{Finn:2000sy, Hughes:1999bq, Hughes:2001jr}. In this work, we assume that the smaller object is in a stable circular orbit (SCO) around the supermassive polymerized black hole at every moment. The properties of a test particle moving along a SCO around the polymerized black hole should satisfy \cite{Misner:1973prb}
\bqn\lb{SCO-condition}
\dot{r}=0,~~~~\f{d V_{\tx{eff}} }{d r} = 0,~~~~ \tx{and}~~~~ \f{d^2 V_{\tx{eff}} }{d r^2} \geq 0.
\eqn
Obviously, the parameter $\hat{k}$ will affect the properties of the SCO in Eq.~\eqref{SCO-condition}. We numerically solve Eq.~\eqref{SCO-condition}, and plot the relations between the radius of the SCO and the parameter $\hat{k}$, and between the energy of the SCO and the parameter $\hat{k}$ in Fig.~\ref{plot-SCO}. One can find that, with a fixed orbital angular momentum, both the radius and the energy of the test particle along the SCO decrease with the parameter $\hat{k}$. 
The innermost stable circular orbit (ISCO) is the SCO with the minimal radius. In gravitational wave astrophysics, the ISCO marks the boundary between the inspiral and merger phases in a two-body system. The properties of a test particle moving along an ISCO around the polymerized black hole should satisfy the conditions: 
\bqn\lb{ISCO-condition}
\dot{r} = 0,~~~~\f{d V_{\tx{eff}} }{d r} = 0,~~~~\f{d^2 V_{\tx{eff}} }{d r^2} = 0.
\eqn
We numerically solve Eq.~\eqref{ISCO-condition}, and plot the relations between the radius of the ISCO and the parameter $\hat{k}$, the orbital angular momentum of the ISCO and the parameter $\hat{k}$, and the energy of the ISCO and the parameter $\hat{k}$ in Fig.~\ref{plot-ISCO}. It shows that all of the radius, the orbital angular momentum, and the energy of the ISCO increase with the parameter $\hat{k}$. 

\section{gravitational waveforms from quasi-circular orbits}\lb{sec3}
\renewcommand{\theequation}{3.\arabic{equation}}
\setcounter{equation}{0} 

In this section, we investigate the orbital evolution of a stellar-mass object inspiraling into a central supermassive polymerized black hole with the quasi-circular orbits approximation and study the corresponding gravitational waveforms. For the smaller object with mass $m$ moving along a trajectory $Z^i(t)$, we treat the Boyer-Lindquist coordinates as a fictitious spherical polar coordinate, then project the object's trajectory  onto the following Cartesian coordinate \cite{gravity-book}
\bqn
x = r \sin \theta \cos \phi,~~~~y = r \sin \theta \sin \phi,~~~~z = r \cos \theta.
\eqn
Then the symmetric and trace-free (STF) mass quadrupole of the smaller object can be defined as \cite{Thorne:1980ru}
\bqn\lb{Iij}
I^{ij} = \left[ \int d^3x T^{tt}(t, x^l) x^i x^j  \right]^{(\text{STF})},
\eqn 
where 
\bqn\lb{Ttt}
T^{tt}(t, x^l) = m \delta^{(3)} [x^l - Z^l(t)].
\eqn
is the $tt$-component of the stress-energy tensor for the object. Gravitational radiation carries away the smaller object's energy and orbital angular momentum. The fluxes of the energy and orbital angular momentum are given by \cite{Maggiore:2007ulw}
\bqn\lb{fluex-1}
\f{dE}{dt} = \f{1}{5} \left< \dddot{I}_{ij} \dddot{I}_{ij} \right>,~~~~\f{dL^i}{dt} = \f{2}{5} \epsilon ^{ikl} \left< \ddot{I}_{ka} \dddot{I}_{la} \right>,
\eqn
where $\epsilon ^{ikl}$ is the three-dimension Levi-Civita symbol and the angle brackets denote the average of a physical quantity over several orbital periods. For a smaller object moving along a circular orbit, Eq.~\eqref{fluex-1} can be simplified to
\bqn\lb{fluex-2}
\f{dE}{dt} = \f{32}{5} m R^4 \Omega^6,~~~~\f{dL^i}{dt} = \f{32}{5} m R^4 \Omega^5,
\eqn
where $R$ is the radius of the circular orbit and $\Omega$ is the angular velocity of the object. 

We assume that the stellar-mass object is in a SCO around the central supermassive polymerized black hole at every moment. Due to gravitational radiation, the energy and orbital angular momentum of the smaller object decrease, leading to a reduction in the radius of the SCO in which the smaller object resides at the same time. Therefore, the complete orbit of the smaller object is quasi-circular. The algorithm we used for modeling the orbital evolution of a stellar-mass object around a supermassive black hole under gravitational radiation proceeds as follows: i) Determine the small object's mass $m$, the supermassive black hole's mass $M$, and a time step $\Delta t$ for evolution. ii) Set the initial position $(r_0, \phi_0)$ of the object. iii) Calculate the initial energy $E_0$, orbital angular momentum $L_0$, and angular velocity $\Omega_0$ of the object. iv) Calculate the mass quadrupole moment $I_{ij}$, energy flux $d E/dt$, and angular momentum flux $d L/dt$ at the object's position. v) Update the object's orbital angular momentum. vi) Compute the new position and energy using the updated orbital angular momentum. vii) Repeat steps iv) through vi), until its orbit shrinks to or below the ISCO. 

\begin{figure}[!t]
	\centering
	\includegraphics[scale =0.80]{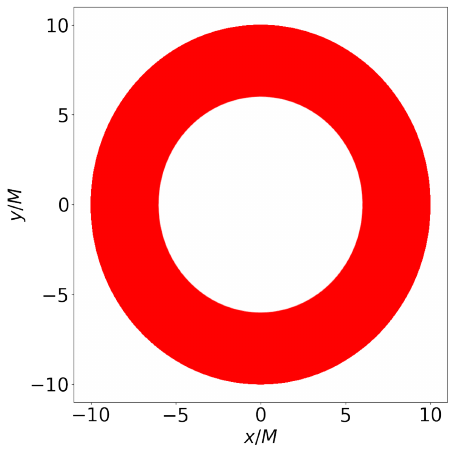}\lb{plot-orbit}
    \caption{The complete quasi-circular orbit of a stellar-mass object inspiraling into a central supermassive polymerized black hole with $\hat{k} = 0.1$. The initial position $(r_0, \phi_0)$ of the object is $(10 M, ~ \pi/2)$.}
	\label{plot-orbit-E}
\end{figure}

\begin{figure}[!t]
	\centering
	\subfigure[]{\includegraphics[scale =0.28]{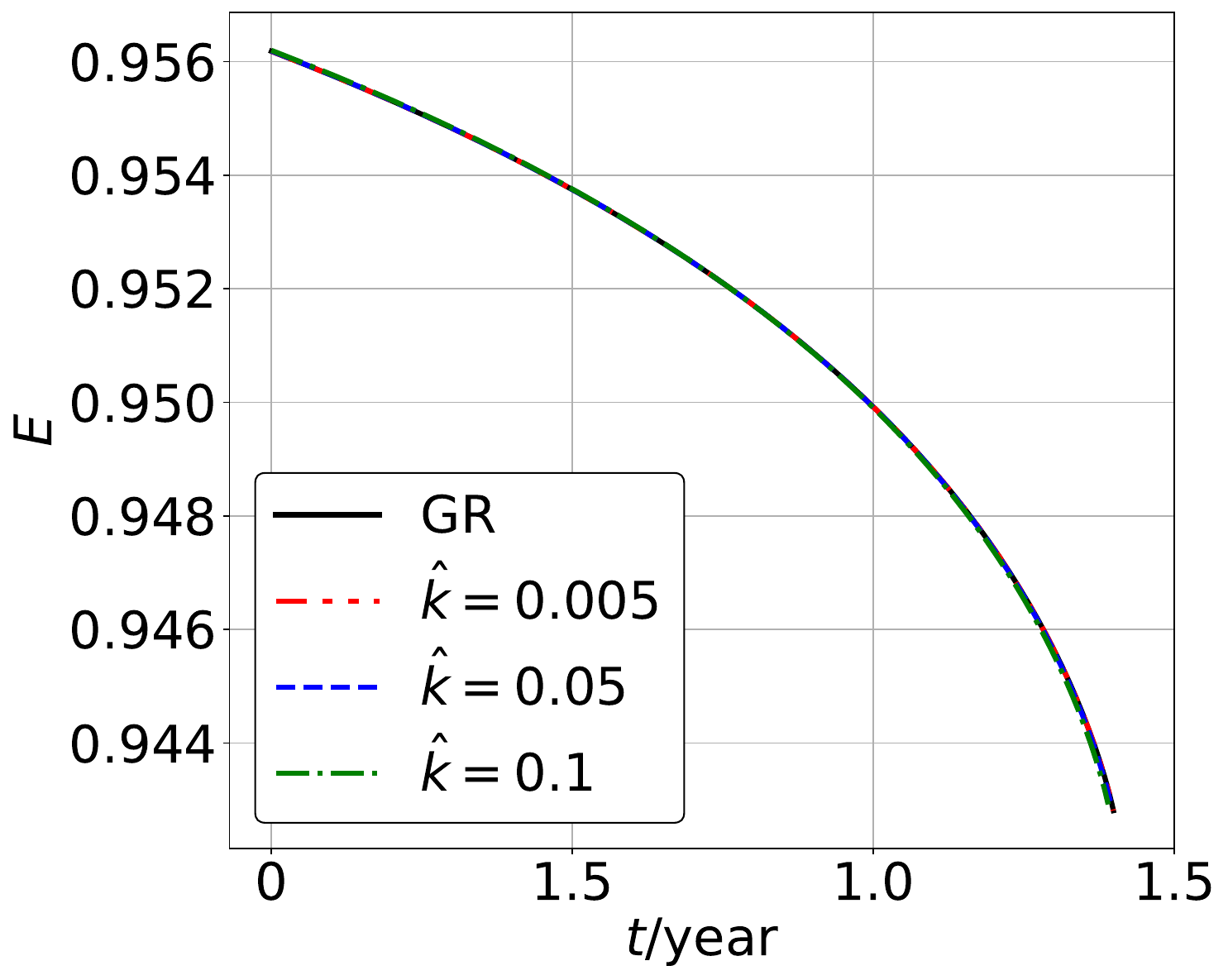}\lb{}}
    \subfigure[]{\includegraphics[scale =0.28]{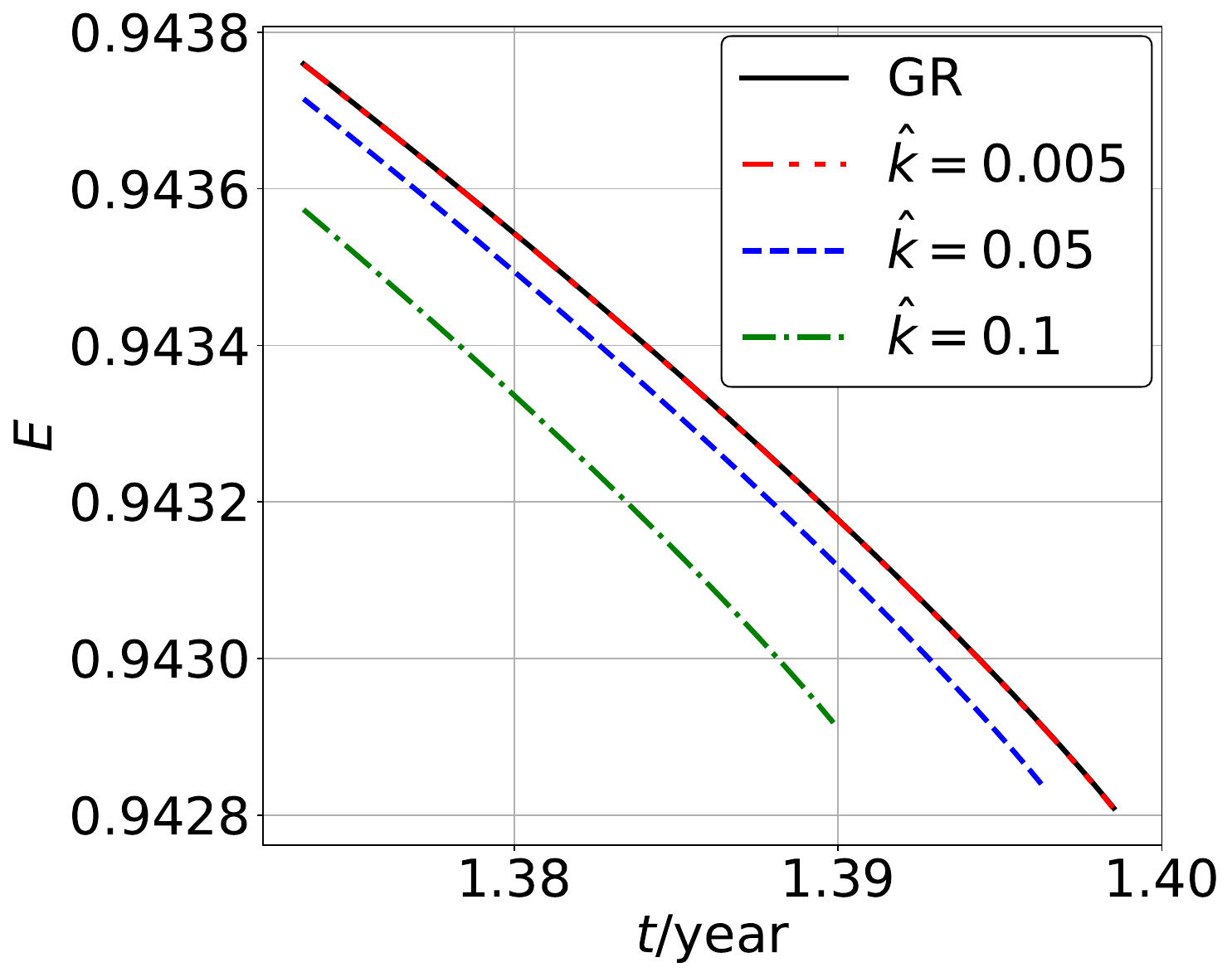}\lb{}}
    \subfigure[]{\includegraphics[scale =0.28]{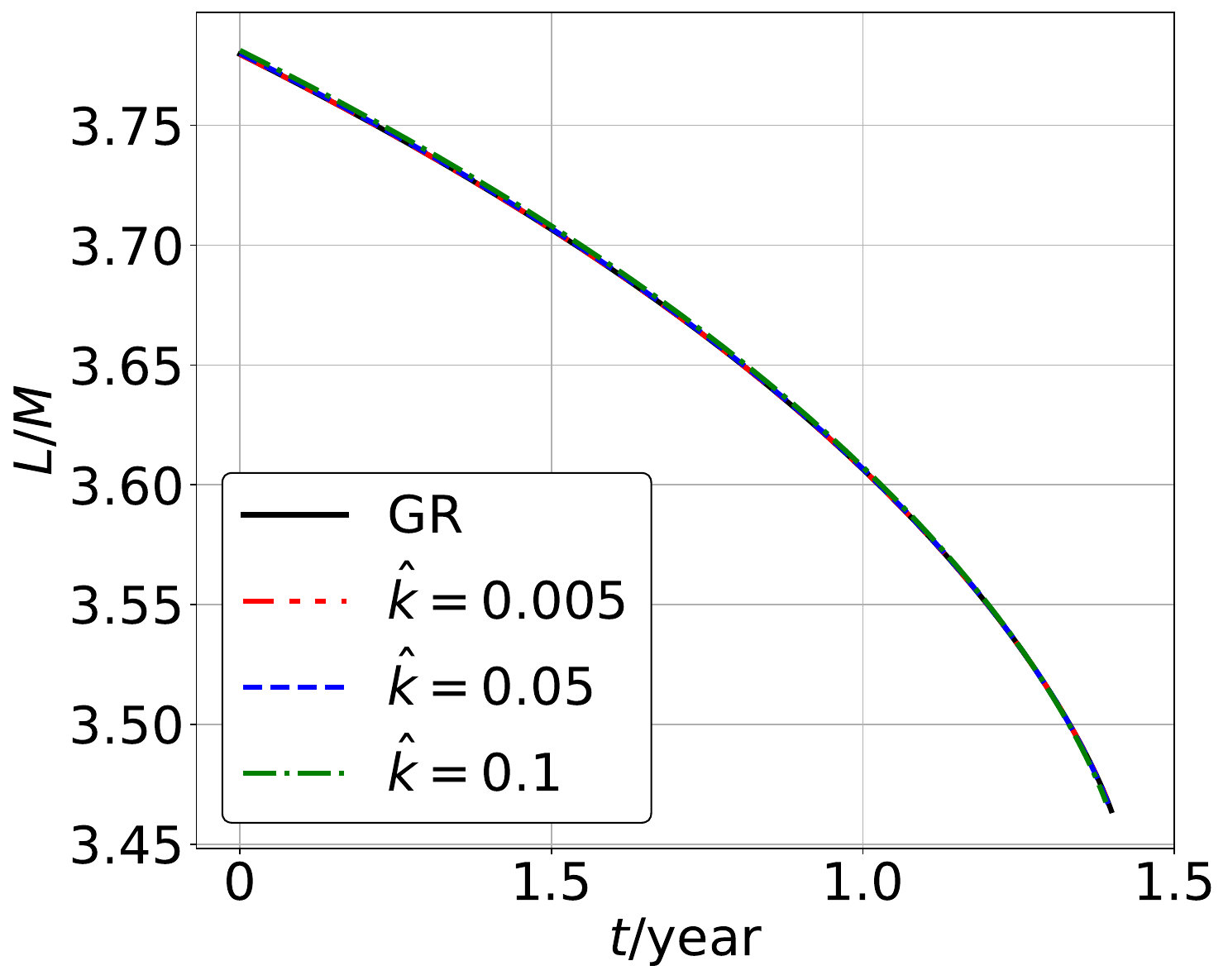}\lb{}}
    \subfigure[]{\includegraphics[scale =0.28]{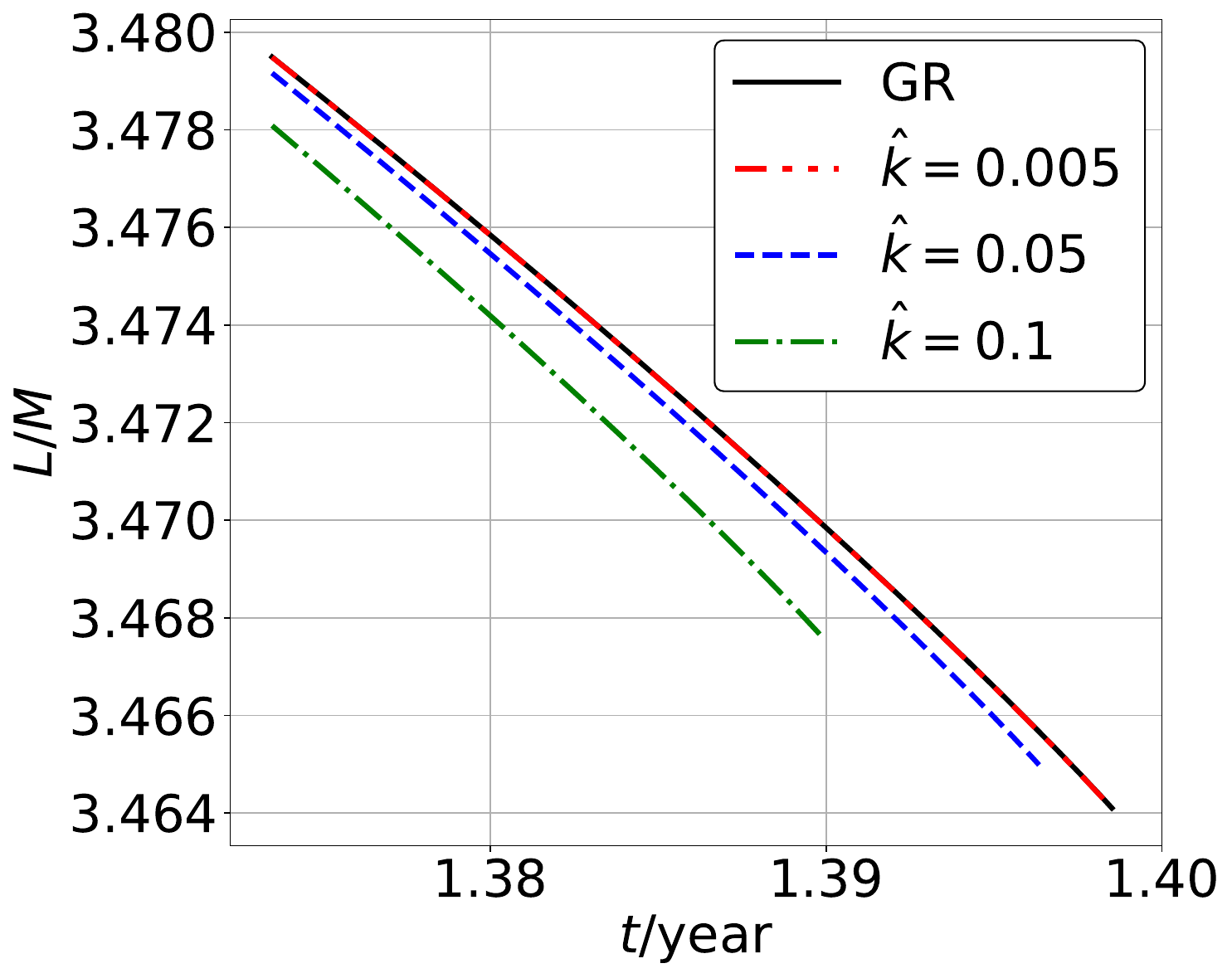}\lb{}}
    \subfigure[]{\includegraphics[scale =0.28]{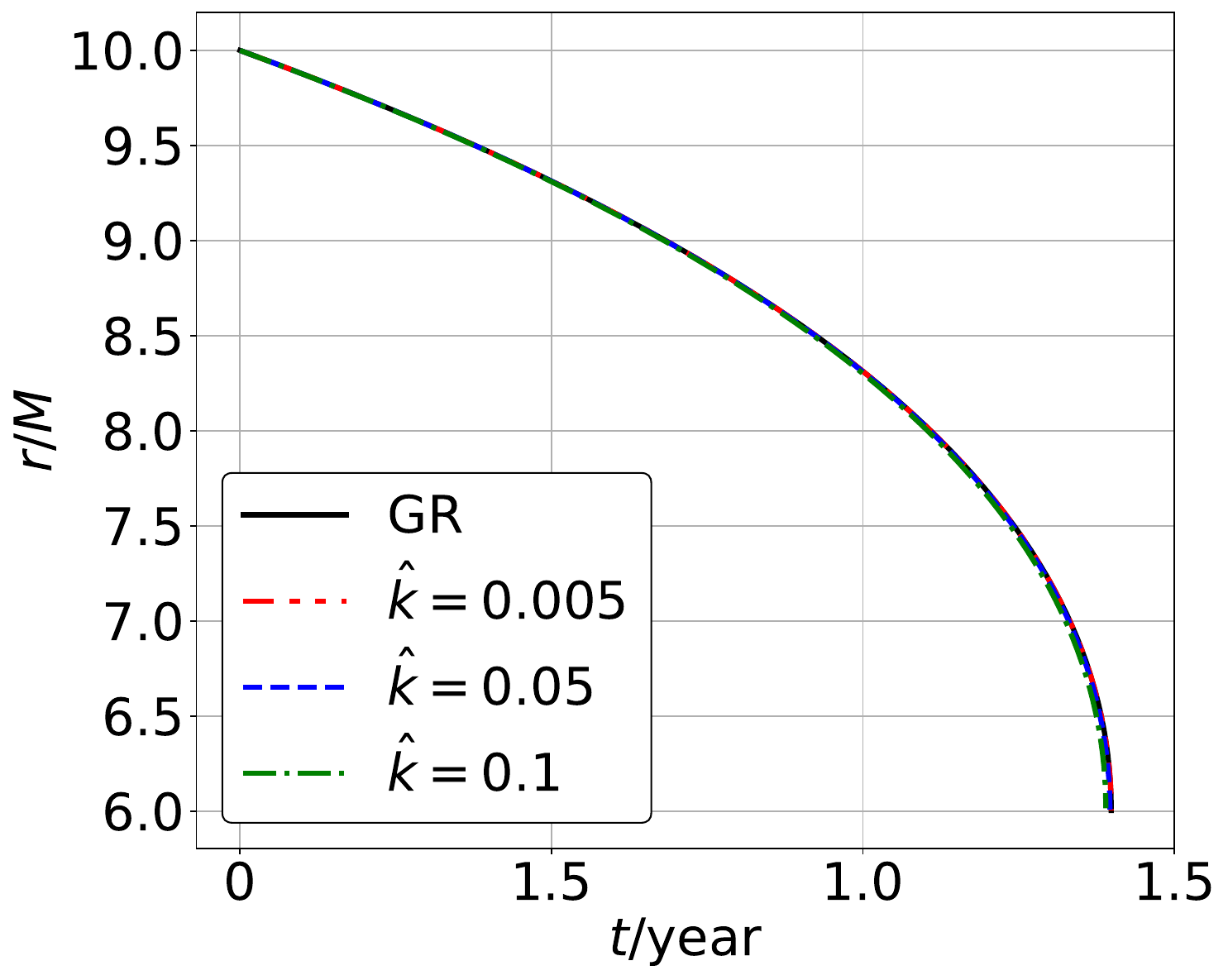}\lb{}}
    \subfigure[]{\includegraphics[scale =0.28]{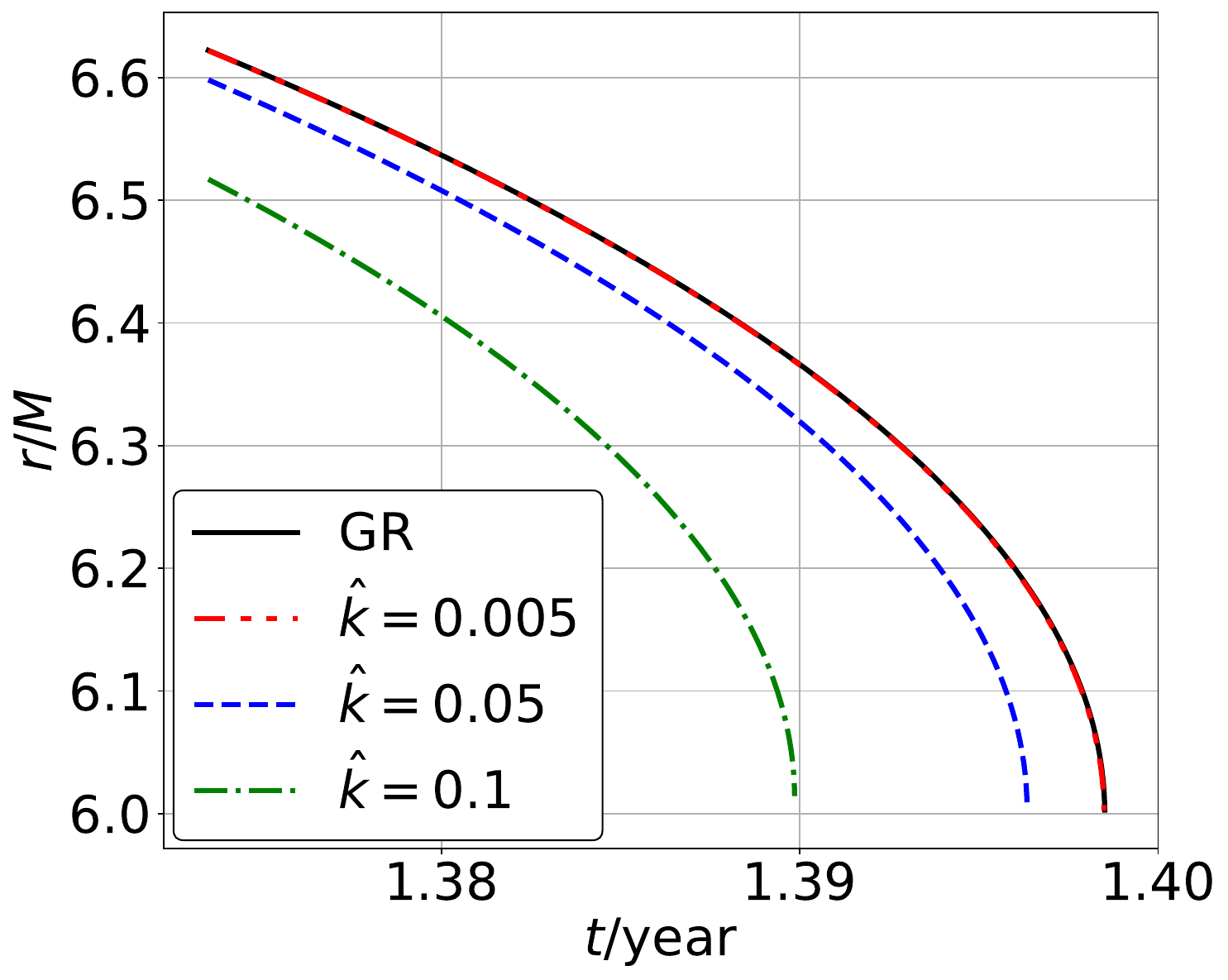}\lb{}}
	\caption{The evolutions of the smaller object's energy $E$, orbital angular momentum $L/M$, and radius $r/M$ of the quasi-circular orbit  around the supermassive polymerized black hole are shown in (a), (c), and (e) within the complete orbital time, and in (b), (d), and (f) within the late stage.}
	\label{Evolutions of ELR}
\end{figure}

The algorithm described above systematically tracks the gradual decay of the orbit due to gravitational radiation, providing a detailed picture of the orbital evolution of a stellar-mass object around a supermassive black hole. Considering an EMRI system consisting of a test object with $m = 10 M_\odot$ and a central supermassive polymerized black hole with $M= 10^6 M_\odot$. By setting the initial position $(r_0, \phi_0)$ of the smaller object as $(10M,~\pi/2)$ and assigning different values to the quantum correction parameter $\hat{k}$, we use the algorithm described above to investigate the evolution of the smaller object's quasi-circular orbit around the central supermassive polymerized black hole. Taking $\hat{k} = 0.1$ as an example, we plot the complete orbit of the smaller object in Fig.~\ref{plot-orbit-E}. We also plot the evolutions of the smaller object's energy $E$, orbital angular momentum $L/M$, and radius $r/M$ of the quasi-circular orbit around the
supermassive polymerized black hole with different values of the parameter $\hat{k}$ in Fig.~\ref{Evolutions of ELR}. The results show that the smaller object's orbital evolution takes a shorter time when it orbits around the central supermassive polymerized black hole with a larger parameter $\hat{k}$. It is consistent with that the radius of the ISCO increases with the parameter $\hat{k}$. We have numerically confirmed the validity of the orbital evolution algorithm we employed in Appendix \ref{AppB}.

After obtaining the orbital evolution of a small object inspiraling into the supermassive polymerized black hole, we can investigate the corresponding gravitational waveforms to further explore the properties of the polymerized black hole. For a stellar-mass object moving around a supermassive black hole, to the order of the mass quadrupole moment, the metric perturbations describing the gravitational waves are defined as \cite{Thorne:1980ru}
\bqn\lb{metric perturbations}
h_{ij} = \f{2}{D_\tx{L}} \frac{d^2 I_{ij}}{dt^2},
\eqn 
where $D_\tx{L}$ is the luminosity distance from the EMRI system to the detector. To obtain the gravitational-wave polarizations, we should construct a new detector-adapted coordinate system $(X,~ Y,~ Z)$~\cite{gravity-book}. Under the detector-adapted coordinate system $(X,~ Y,~ Z)$, the gravitational-wave polarizations from Eq.~\eqref{metric perturbations} are
\bqn
h_{+} &=& (h_{\zeta \zeta} - h_{\iota \iota})/2, \lb{waveform-p}\\
h_{\times} &=& h_{\iota \zeta}.\lb{waveform-c}
\eqn
The components $h_{\zeta \zeta}$, $h_{\iota \iota}$,  and $h_{\iota \zeta}$ are given by  \cite{Babak:2006uv}
\bqn
h_{\zeta \zeta} &=& h_{xx} \cos^2 \zeta - h_{xy}  \sin 2 \zeta + h_{yy} \sin^2 \zeta,\\
h_{\iota \iota} &=&  \cos^2 \iota [h_{xx} \sin^2 \zeta + h_{xy} \sin 2 \zeta + h_{yy} \cos^2 \zeta] + h_{zz} \sin^2 \iota - \sin 2 \iota [h_{xz} \sin \zeta + h_{yz} \cos \zeta], \\
h_{\iota \zeta} &=& \cos \iota \left[ 
 \f{1}{2} h_{xx} \sin 2 \zeta + h_{xy} \cos 2 \zeta - \f{1}{2} h_{yy} \sin 2 \zeta \right] + \sin \iota [h_{yz} \sin \zeta - h_{xz} \cos \zeta], 
\eqn 
where $\iota$ is the inclination angle of the orbital plane of the smaller object to the $X-Y$ plane and $\zeta$ is the longitude of the pericenter measured in the orbital plane.

\begin{figure}[!t]
	\centering
	\subfigure[]{\includegraphics[scale =0.23]{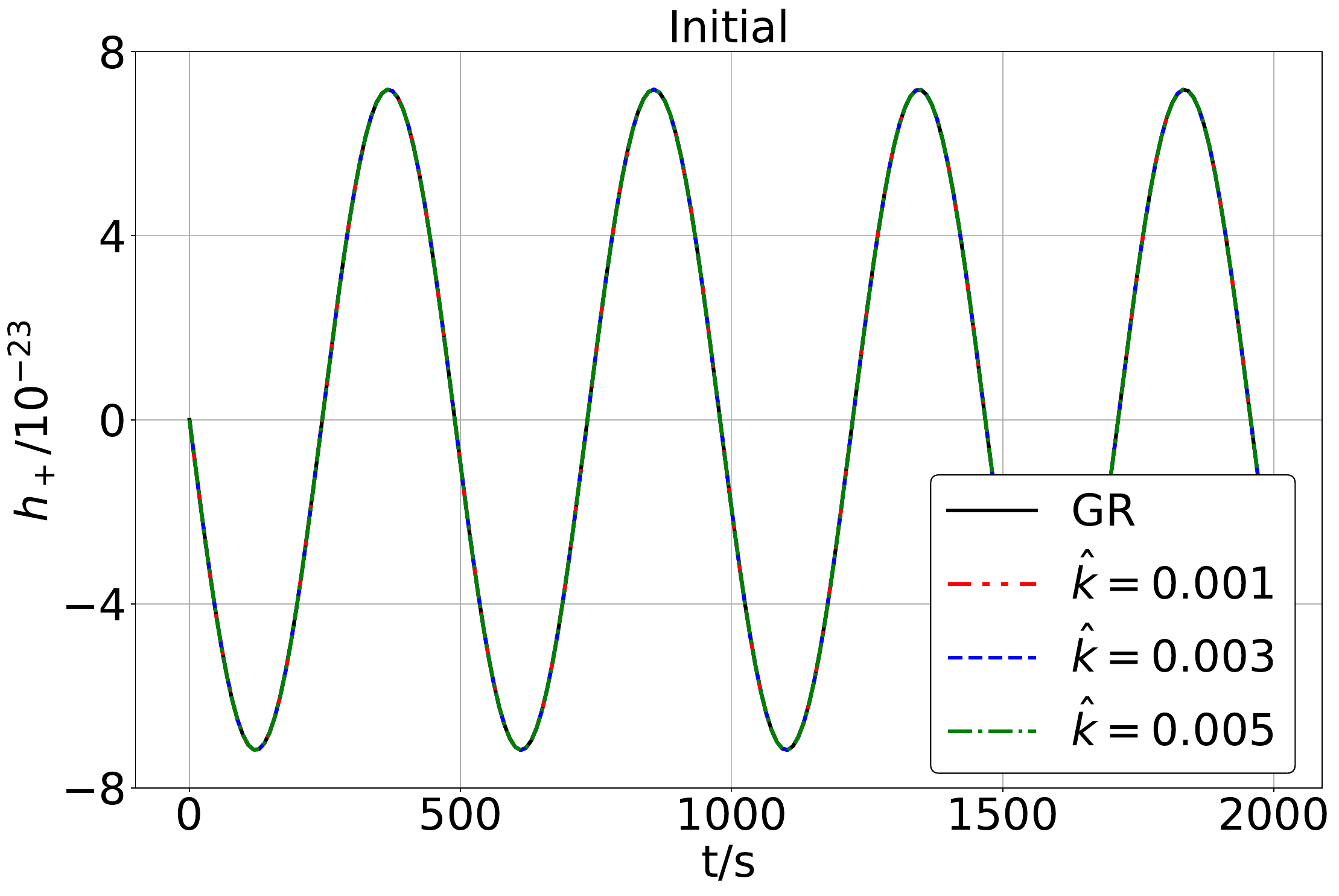}\lb{hp_initial}}
    \subfigure[]{\includegraphics[scale =0.23]{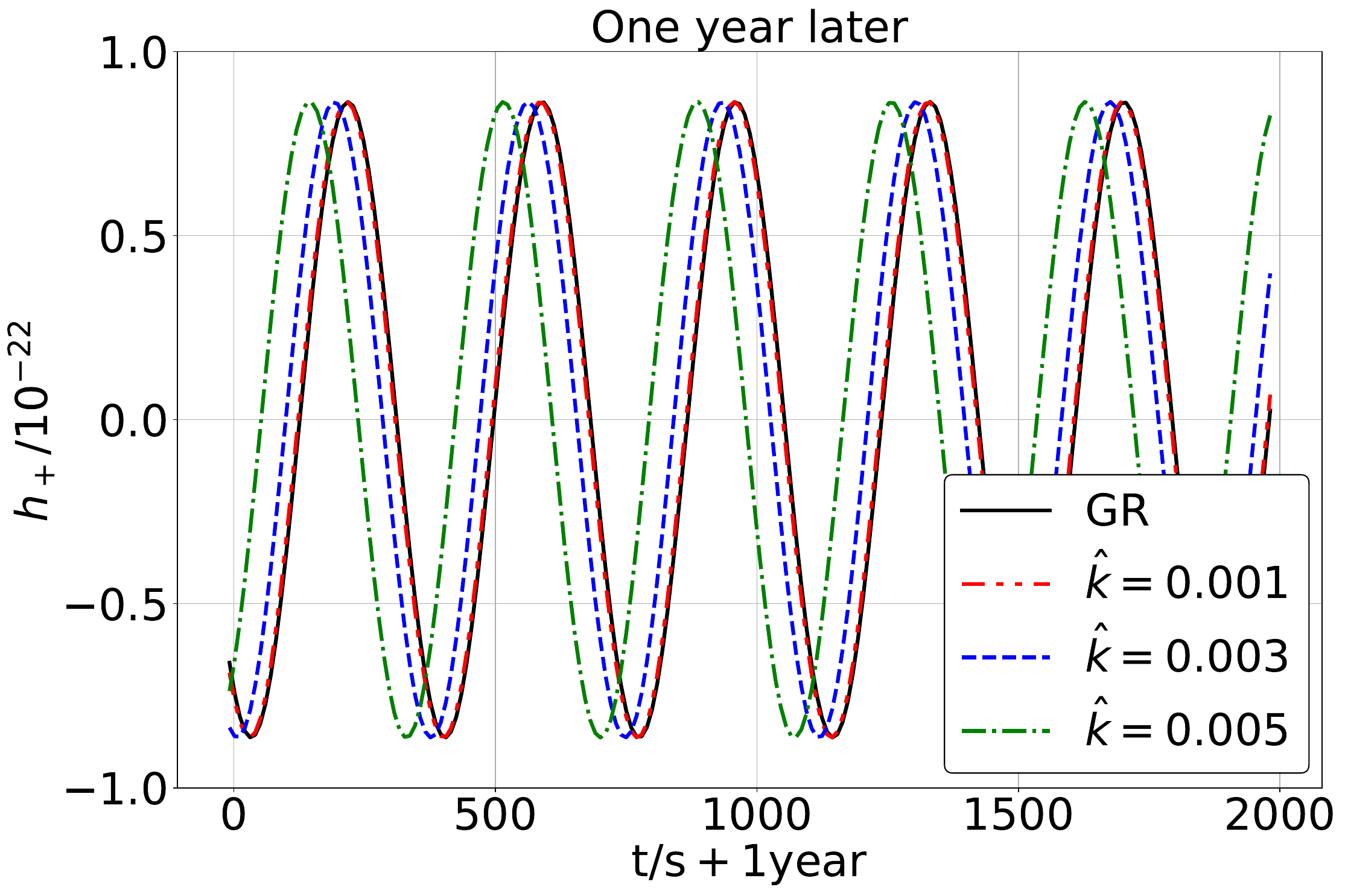}\lb{hp_late}}
    \subfigure[]{\includegraphics[scale =0.23]{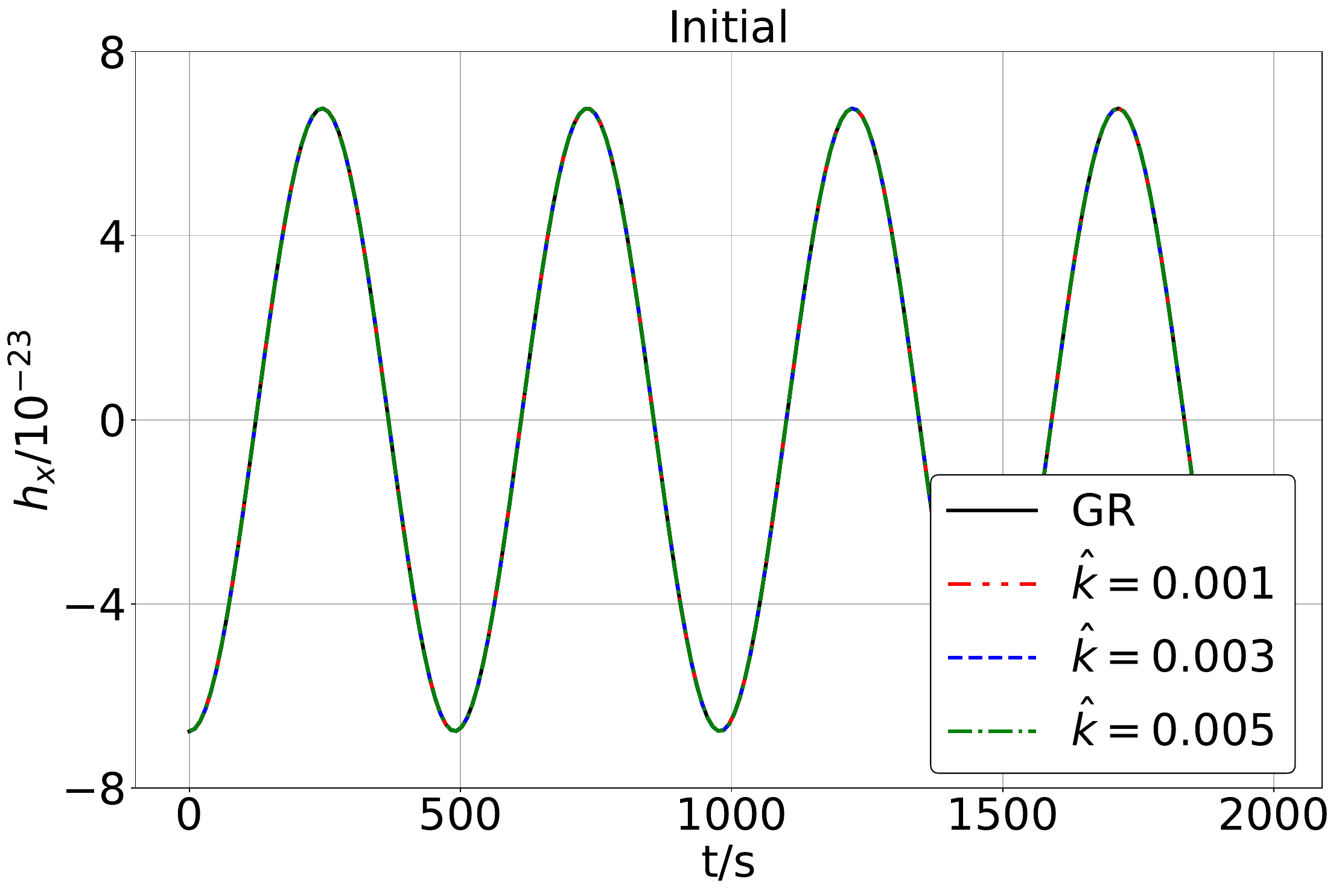}\lb{hc_initial}}
    \subfigure[]{\includegraphics[scale =0.23]{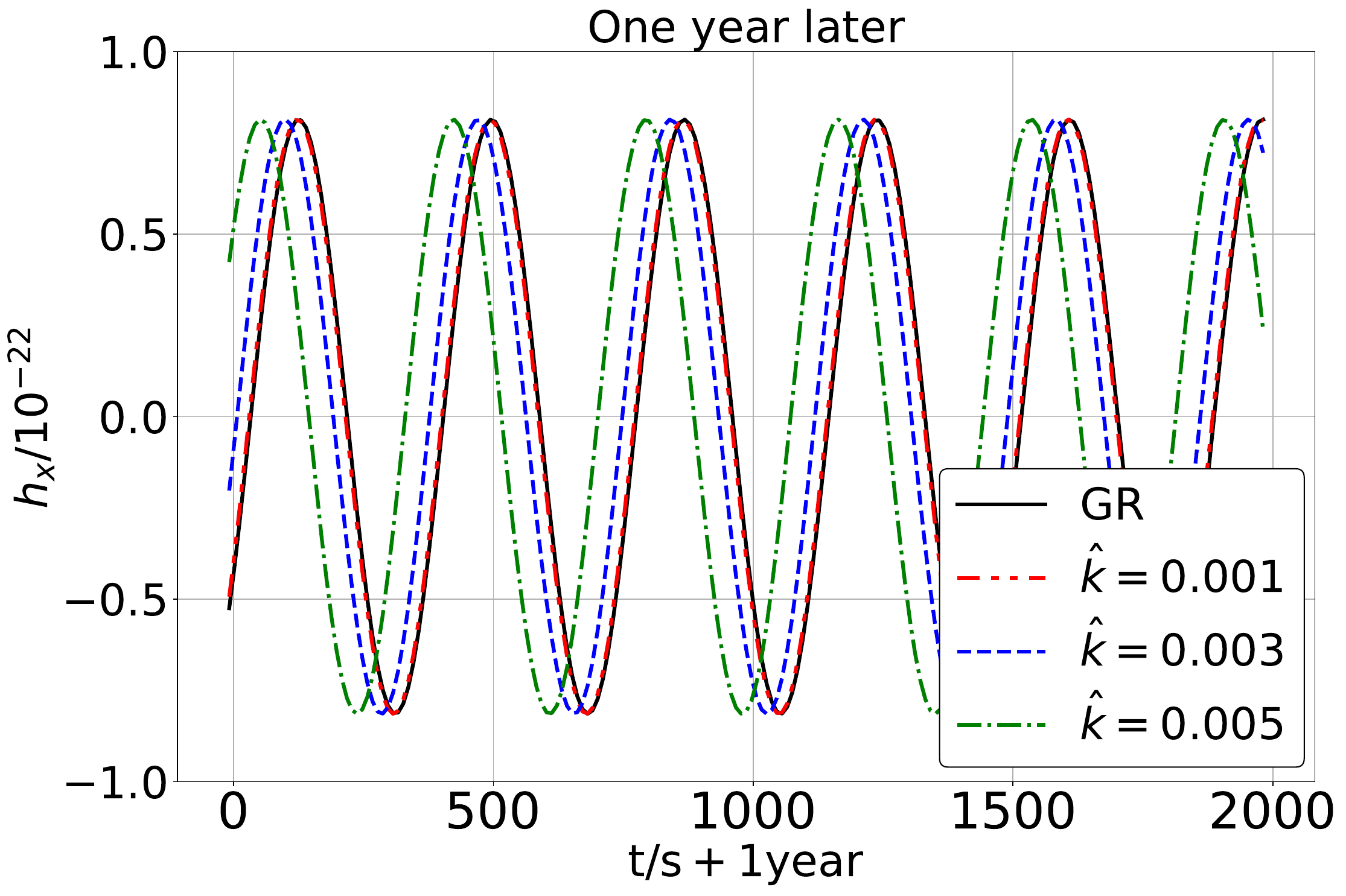}\lb{hc_late}}
	\caption{Gravitational waveforms from a test object with $m = 10 M_\odot$ along quasi-circular orbits around a supermassive polymerized black hole with $M = 10^6 M_\odot$ and different values of the parameter $\hat{k}$. (a) and (c): The gravitational waveforms within the initial 2000 seconds. (b) and (d): The gravitational waveforms within 2000 seconds one year later.}
	\label{gravitational waveforms}
\end{figure}

Throughout the evolution of the smaller object's quasi-circular orbit, we get the smaller object's mass quadrupole moment at each position. Then, we calculate the corresponding gravitational waveforms from Eqs. \eqref{waveform-p} and \eqref{waveform-c}, with $D_\text{L} = 2 \text{Gpc}$, $\iota = \pi/4$, and $\zeta = \pi/4$. We plot the gravitational waveforms in Fig.~\ref{gravitational waveforms}. Figures \ref{hp_initial} and \ref{hc_initial} show that the parameter $\hat{k}$ has almost no effect on the gravitational waveforms during the initial 2000 seconds. And one can find from Figs. \ref{hp_late} and \ref{hc_late} that the impact of the parameter $\hat{k}$ becomes visually apparent after a
one-year accumulation and the parameter $\hat{k}$ causes an advance in the phase of the gravitational waveforms.

\section{dephasing and mismatch}\lb{sec4}
\renewcommand{\theequation}{4.\arabic{equation}}
\setcounter{equation}{0} 

In the study of gravitational waves from EMRIs around black holes in modified gravity theories, the impact of modified gravity on the phase of the gravitational waves is more pronounced than its effect on the amplitude. This is because the phase is closely linked to the orbital dynamics and spacetime geometry. Changes in the gravitational potential due to modified gravity can cause shifts in the orbital frequency, which in turn modify the phase evolution over time. While the amplitude of the gravitational waves also carries valuable information, the phase typically encodes the most detailed signatures of the underlying gravitational theory. As a result, analyzing the phase evolution offers a more sensitive probe for detecting the effects of modified gravity in EMRIs.

\begin{figure}[!t]
	\centering
	\includegraphics[scale =0.28]{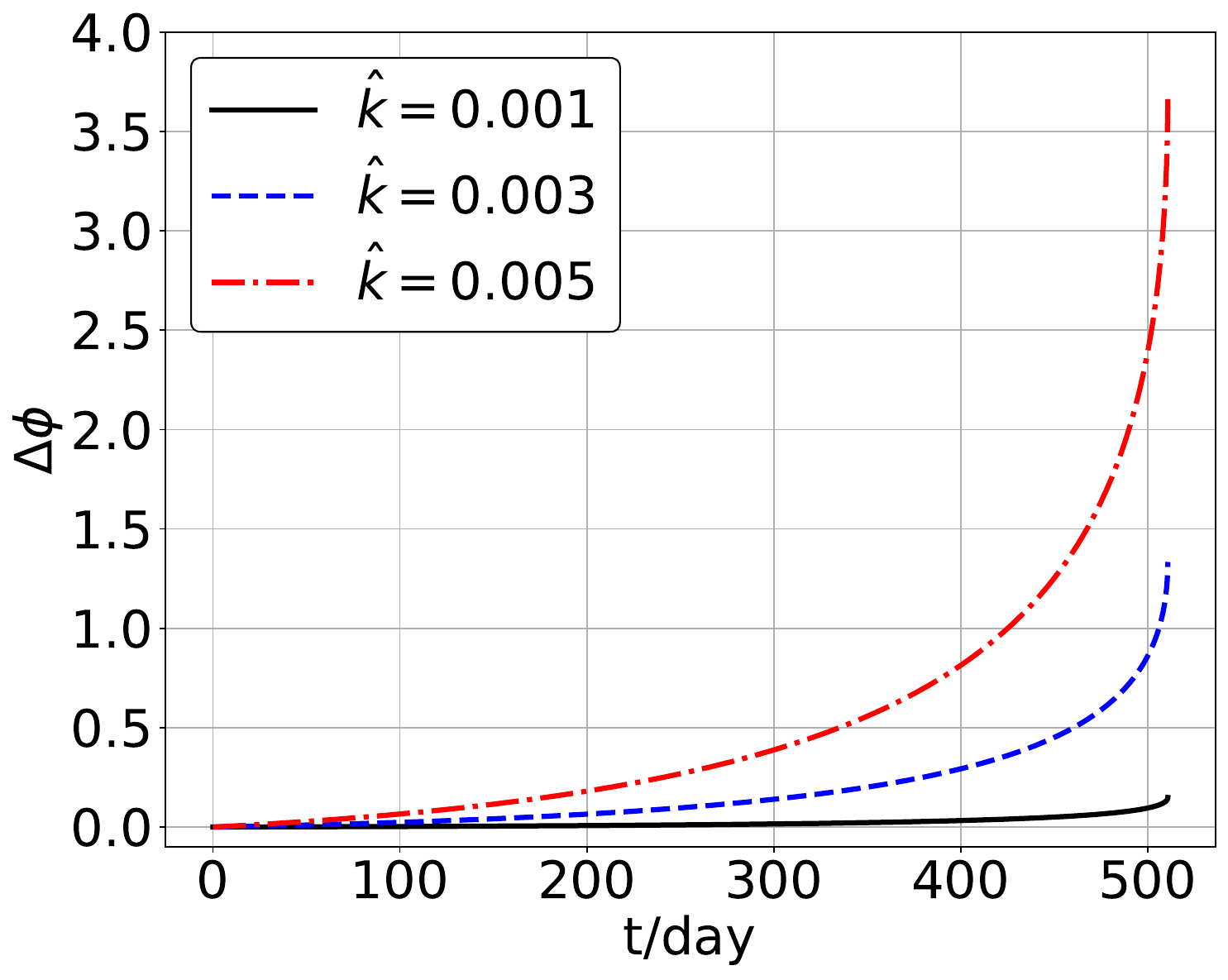}
	\caption{The dephasing of the gravitational waves with different values of the parameter $\hat{k}$.}
	\label{phase}
\end{figure}

To further investigate the effect of the parameter $\hat{k}$ on the gravitational waves' phase, we define the dephasing for the gravitational waves with $\hat{k}$ as
\bqn\lb{phase-Eq}
\Delta \Phi = \Phi(\hat{k}) -\Phi_0, 
\eqn 
where $\Phi_0$ is the phase of the gravitational waves in the supermassive Schwarzschild black hole case. We calculate the dephasing of the gravitational waves with different values of the parameter $\hat{k}$ though Eq.~\eqref{phase-Eq}. The results are shown in Fig. \ref{phase}. One can find that the parameter $\hat{k}$ induces a phase advance in the gravitational waveforms, with a larger parameter resulting in a more pronounced phase shift. And this phase advancement accumulates over time.

To explore the observability of the quantum parameter's effect on gravitational waves, we should consider the response signal of the gravitational waves in a detector. The response signal of gravitational waves in a detector is related to the characteristics of that detector. The strain amplitude in LISA can be described as \cite{Cutler:1997ta}
\bqn\lb{strain}
h_{\tx{I,II}} = \f{\sqrt{3}}{2}  \left( F^{+}_{\tx{I,II}} h_{+}+ F^{\times}_{\tx{I,II}} h_{\times} \right),
\eqn 
where the factor $\sqrt{3}/2$ is from the fact that the actual angle between LISA arms is $60^\circ$, and $F^{+, \times}_{\tx{I,II}}$ are the ``antenna pattern'' functions. Here we use the expressions of the ``antenna pattern'' functions in Ref.~\cite{Barack:2003fp}. The ``antenna pattern'' functions depend on both the direction of the source $(\theta_{S},~ \phi_{S})$ and the orbital angular momentum direction $(\theta_{L} = 0,~\phi_{L} = 0)$. The motion of LISA introduces a Doppler shift correction to the phase of the gravitational waves. Following Ref.~\cite{Barack:2003fp}, we modify the phase of the gravitational waves as \bqn
\Phi(t)_{\text{corrected}} = \Phi(t) + \Phi(t)' R_\tx{AU} \sin \theta_\tx{S} \cos\left( 2 \pi t/T - \phi_\tx{S} \right),
\eqn 
where $R_\tx{AU}$ is the astronomical unit and $T = 1$ year.

For two response signals, $h_1(t)$ and $h_2(t)$, the noise-weighted inner product between those is defined as \cite{Flanagan:1997sx, Lindblom:2008cm}
\bqn
<h_1|h_2> = 2 \sum_{\lambda = \text{I,II}}\int^{f_\tx{max}} _{f_\tx{min}} \f{\tilde{h}_{1,\lambda}(f) \tilde{h}^\ast_{2,\lambda}(f) + \tilde{h}^\ast_{1,\lambda}(f) \tilde{h}_{2,\lambda}(f) }{S_{n}(f)} df,
\eqn
where $\tilde{h}_i(f)$ is the Fourier transformation of $\tilde{h}_i(t)$, $\tilde{h}^\ast_i(f)$ is the complex conjugate of $\tilde{h}_i(t)$, $f_\tx{min}$ and $f_\tx{max}$ are the boundaries of the frequency range of both $\tilde{h}_1(f)$ and $\tilde{h}_2(f)$, and $S_n(f)$ is the power spectral density of LISA \cite{Larson:1999we, Robson:2018ifk}. The signal-to-noise ratio (SNR) of a response signal $h$ is $\sqrt{<h|h>}$. To evaluate the degree of similarity between gravitational wave signals $h_1$ and $h_2$, we introduce a faithfulness function as \cite{Lindblom:2008cm}
\bqn\lb{faithfulness function}
\mathcal{F}[h_1, h_2] = \mathop{\text{max}}\limits_{\{t_C,\phi_C\}} \frac{<h_1|h_2>}{\sqrt{<h_1|h_1><h_2|h_2>}}.
\eqn
The corresponding mismatch function is \cite{Lindblom:2008cm} 
\bqn
\mathcal{M}[h_1, h_2] = 1 - \mathcal{F}[h_1, h_2].
\eqn
For the gravitational wave signals with SNR $\rho$ and from a model with $N$ intrinsic parameters, the critical value of the mismatch is $N/(2 \rho^2)$~\cite{Lindblom:2008cm}. If the mismatch between two gravitational wave signals is smaller than the critical mismatch, these two signals are indistinguishable by the detector. 

\begin{figure}[!t]
	\centering
	\includegraphics[scale =0.28]{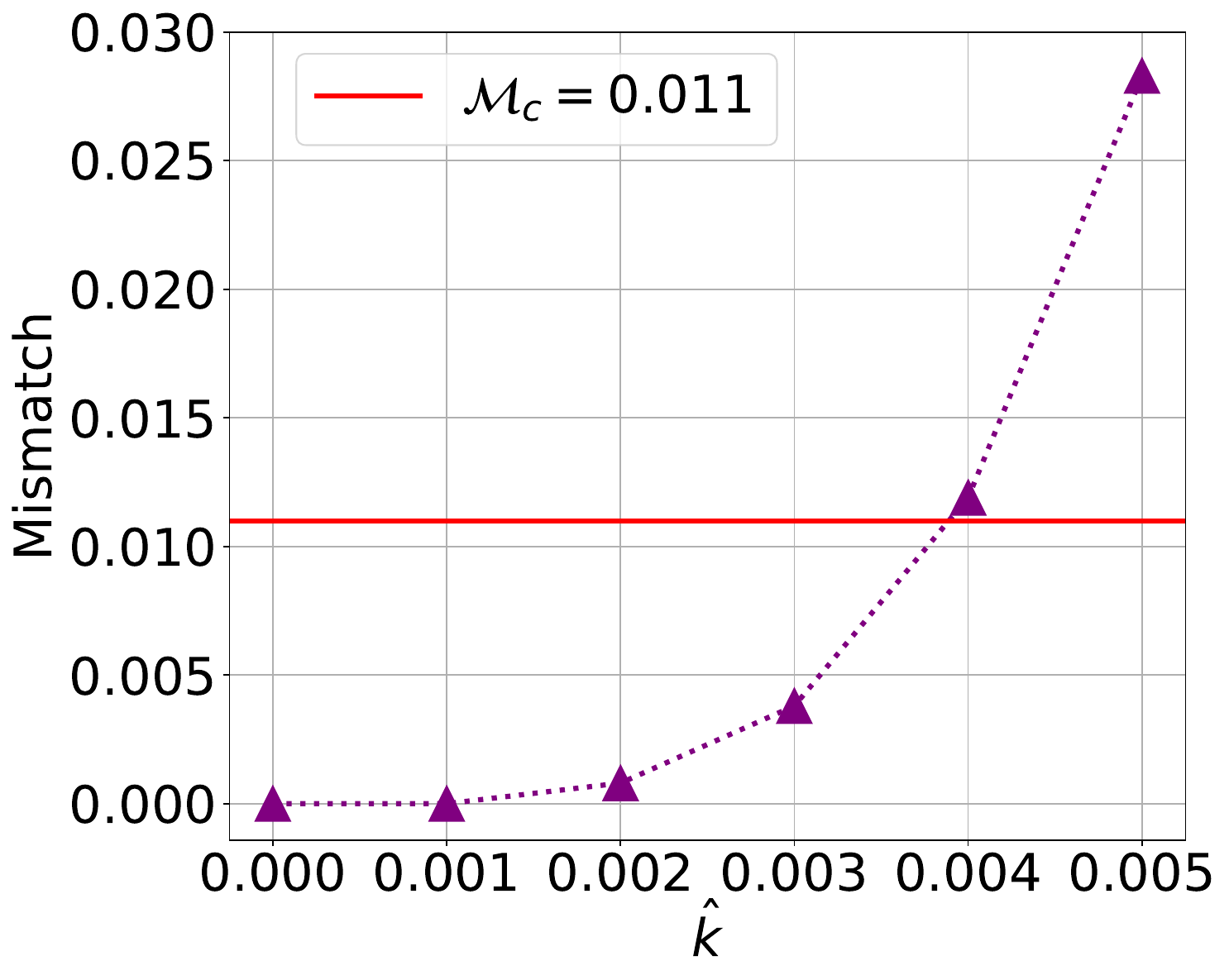}
	\caption{Mismatch between the gravitational waves with different values of the parameter $\hat{k}$ and that for the supermassive Schwarzschild black hole.}
	\label{Mismatch}
\end{figure}

For the model of the gravitational waveforms in this work, the dimension of the intrinsic parameters $N=6$. In fact, there are two strain signals for a gravitational wave detected by LISA, as Eq.~\eqref{strain}. So, the final SNR of a gravitational wave is $\rho_\text{F} = \sqrt{\rho^2_\text{I} + \rho^2_\text{II}}$.
Following Refs.~\cite{Maselli:2021men,Barsanti:2022vvl}, we consider
the last one-year evolution of the EMRI system and rescale the luminosity distance $D_\text{L}$ to make the final SNR $\rho_\text{F}$ to be
30. Then, the corresponding  threshold value for the final mismatch is $0.0033$. We calculate the mismatch between the gravitational waves for the supermassive polymerized black hole with different values of the parameter $\hat{k}$ and that for the supermassive Schwarzschild black hole. The results are shown in Fig.~\ref{Mismatch}. One can find that, under setting $\rho_\text{F}=30$, the mismatch between the gravitational waves for the supermassive polymerized black hole with $\hat{k} < 0.003$ and that for the supermassive Schwarzschild black hole is smaller than the critical mismatch. It means that we can probe the parameter $\hat{k}$ to $\mathcal{O}(10^{-3})$, which is tighter than the constraint $\hat{k} < 0.36$ from the Sgr A$^\ast$ black hole shadow \cite{KumarWalia:2022ddq}, with a one-year observation of the EMRI system.

\begin{figure}[!t]
	\centering
	\includegraphics[scale =0.3]{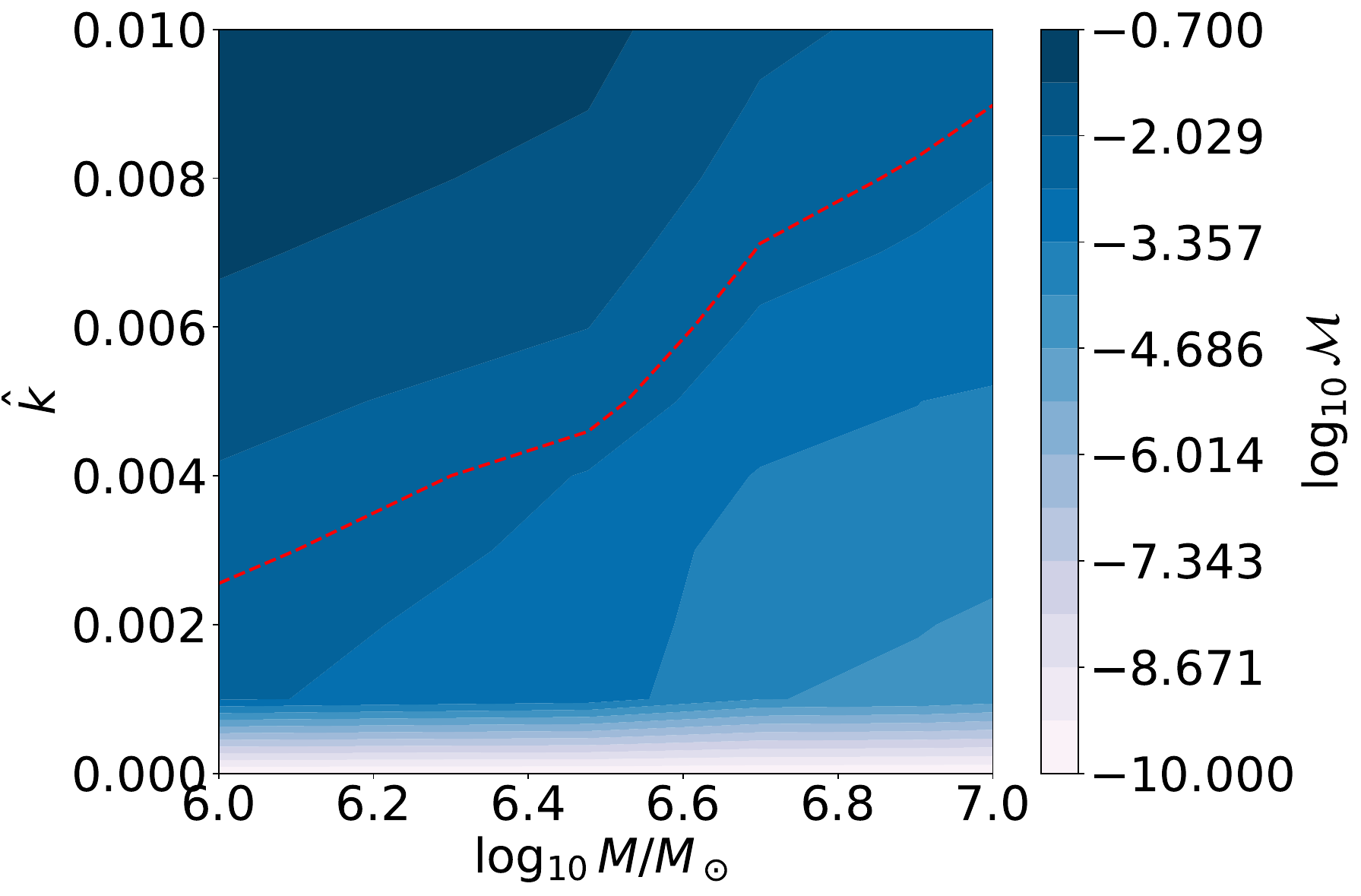}
	\caption{Contour plot of the logarithmic mismatch 
	$\log_{10} \mathcal{M}$
	as a function of the supermassive black hole mass $M$ and the quantum correction parameter $\hat{k}$. The red dashed line indicates the logarithmic critical mismatch trajectory $\log_{10} \mathcal{M}_c = -2.482$ across varying black hole masses.}
	\label{plot-contour}
\end{figure}

To ensure the robustness of the constraints on the quantum correction parameter $\hat{k}$, it is essential to examine potential parameter degeneracies. In particular, since both the black hole mass $M$ and the quantum parameter $\hat{k}$ influence the gravitational waveform, one must verify that their effects are distinguishable through mismatch analysis. If $\hat{k}$ was strongly degenerated with $M$, then the observed waveform could not uniquely determine the presence of quantum corrections. To address this, we compute the logarithmic mismatch $\log_{10} \mathcal{M}$ over a grid of 
$(\log_{10}M/M_\odot,~\hat{k})$ values and present the result in Fig.~\ref{plot-contour}. The color map shows $\log_{10} \mathcal{M}$, with darker shades corresponding to larger mismatches. The red dashed curve traces the logarithmic critical mismatch value at each black hole mass. As Fig.~\ref{plot-contour} shows, the location of the logarithmic critical mismatch $\log_{10} \mathcal{M}_c = -2.482$ varies non-trivially with $M$, and the mismatch exhibits a clear and continuous dependence on both parameters. This behavior indicates that the effect of the quantum correction parameter $\hat{k}$ cannot be fully absorbed by rescaling the black hole mass, thereby demonstrating that $\hat{k}$ and $M$ are not degenerate. This supports the validity of our approach in constraining $\hat{k}$ using gravitational wave signals from EMRIs.

\section{Conclusions and discussions}\lb{sec5}

In this work, we explored the quasi-circular EMRIs around a central supermassive polymerized black hole. By modeling the smaller object as a massive test particle, we derived its equations of motion and the associated radial effective potential. Our analysis revealed that the peak of the radial effective potential decreases with the dimensionless quantum-corrected parameter $\hat{k}$. We examined the properties of SCOs around the polymerized black hole, finding that both the radius and the energy of the particle on these orbits decrease with the parameter $\hat{k}$, given a fixed orbital angular
momentum. Conversely, for a
fixed $\hat{k}$, both the radius and the energy of the particle on the SCOs increase with the particle’s orbital angular momentum. Notably, we also investigated the ISCOs around the polymerized black hole and found that the radius, the orbital angular momentum, and the energy of the ISCOs all increase with the parameter $\hat{k}$.

Next, we adopted a numerical algorithm to evolve the quasi-circular orbits of a small object inspiraling into a supermassive polymerized black hole. Using the numerical kludge scheme, we analyzed the corresponding gravitational waveforms of this quasi-circular EMRI system. We found that the impact of quantum corrections on the gravitational waveforms is initially negligible but becomes visually apparent after one year. Additionally, we examined the influence of the parameter $\hat{k}$ on the phase of the gravitational waveforms, discovering that $\hat{k}$ induces a phase advance that accumulates over time.

Finally, we took into account the Doppler shift caused by LISA’s motion and modified the phase of the gravitational waves accordingly. We then calculated the gravitational wave response signal detected by LISA. To assess the observability of the quantum parameter’s impact on gravitational waves, we computed the mismatch between waveforms generated with different values of the quantum parameter and those corresponding to a supermassive Schwarzschild black hole. Considering the last one-year evolution of the EMRI system and rescaling the luminosity distance to make the SNR 30, we obtained the critical value for the mismatch. Our results indicate that the mismatch between the gravitational waves for the supermassive polymerized black hole with $\hat{k} < 0.003$ and those for the supermassive Schwarzschild black hole
is smaller than the critical value. This suggests that LISA could probe the parameter $\hat{k}$ to $\mathcal{O}(10^{-3})$ with a one-year observation of the EMRI system. It is tighter than the current limit of $\mathcal{O}(10^{-1})$ obtained from the Sgr A$^\ast$ black hole shadow \cite{KumarWalia:2022ddq}. To assess the robustness of our constraints, we investigated the potential degeneracy between the quantum correction parameter $\hat{k}$ and the black hole mass $M$. Our mismatch analysis shows that the effects of $\hat{k}$ cannot be mimicked by adjusting $M$. It indicates that the effects of the two parameters are distinct and lead to non-degenerate states. This supports the reliability of using EMRI waveforms to constrain quantum corrections in the black hole spacetime. Exploring the constraints on the quantum correction parameter through various astronomical observations contributes to the study of quantum-corrected black hole models.  

It is important to note that we employed the quasi-circular inspiral approximation to model the orbital evolution of smaller objects in EMRIs. However, for a more comprehensive analysis, the evolution of elliptical orbits should also be considered \cite{LISA:2022yao}. Additionally, given that celestial objects generally possess spin, incorporating the effects of spin on orbital evolution is essential \cite{Apostolatos:1994mx, Zhang:2018omr}. Conversely, the presence of dark matter surrounding a supermassive black hole can influence both the orbital dynamics of smaller objects and the resulting gravitational waveforms \cite{Barausse:2014tra, Hannuksela:2019vip, Li:2021pxf}. A major challenge lies in distinguishing the effects of modified gravity from those of environmental factors on EMRIs. We will attempt to address these issues in our future work.

\section*{Acknowledgements}

We thank Rui Niu for important discussions. This work was supported by the National Key Research and Development Program of China (Grant No.~2021YFC2203003), the National Natural Science Foundation of China (Grants No.~12475056, No.~12105126, No.~12275238, and No.~12247101), the National Natural Science Foundation of Gansu Province (Grant No.~22JR5RA389), the `111 Center' under Grant No.~B20063, the Zhejiang Provincial Natural Science Foundation of China under Grant No.~LR21A050001 and No.~LY20A050002, and the Fundamental Research Funds for the Provincial Universities of Zhejiang in China under Grant No.~RF-A2019015. 

\appendix

\textcolor{black}{
\section{Motion of the test particle on the equatorial plane}\lb{AppA}}
\renewcommand{\theequation}{A.\arabic{equation}}
\setcounter{equation}{0}
\textcolor{black}{
The Lagrangian of the test particle with mass $m$
is \cite{Chandrasekhar:1985kt}
\bqn\lb{Lagrangian}
\mathscr{L} = \f{m}{2} g_{\mu \nu} \f{d x^\mu}{d \tau} \f{d x^\nu}{d \tau} = -\f{m}{2},
\eqn 
where $\tau$ is the proper time. For simplicity, we set the mass of the test particle $m = 1$. The generalized momentum of the test particle is defined by
\bqn\lb{momentum}
p_{\mu} = \f{\p \mathscr{L}}{\p \dot{x}^\mu} = g_{\mu \nu} \dot{x}^\nu,
\eqn
where the dot denotes the derivative with respect to the proper time. Substituting Eqs.~\eqref{metric} and~\eqref{Lagrangian} into Eq.~\eqref{momentum},
the equations of motion for the test particle are derived as
\bqn
p_{t} &=&  - \left( \sqrt{1- \frac{k^2}{r^2}} - \frac{2M}{r} \right) \dot{t} = - E, \lb{momentum-1} \\
p_{\phi} &=& r^2 \sin^2 \theta \dot{\phi} = L, \lb{momentum-2} \\
p_{r} &=& \lf( \left( \sqrt{1- \frac{k^2}{r^2}} - \frac{2M}{r} \right) \left( 1- \frac{k^2}{r^2} \right)\rt)^{-1} \dot{r} , \lb{momentum-3} \\
p_{\theta} &=& r^2 \dot{\theta} , \lb{momentum-4}
\eqn 
where $E$ and $L$ represent the energy and the orbital angular momentum of the test particle per unit mass, respectively. Here we only consider the test particle moving on the equatorial plane $(\theta = \pi/2)$ around the polymerized black hole. Then from Eq.~\eqref{Lagrangian} we obtain
\bqn\lb{Lagrangian-2}
\f{\dot{r}^2}{\left( \sqrt{1- \frac{k^2}{r^2}} - \frac{2M}{r} \right) \left( 1- \frac{k^2}{r^2} \right)} + \f{L^2}{ r^2 } - \f{E^2}{ \sqrt{1- \frac{k^2}{r^2}} - \frac{2M}{r}} = -1.
\eqn 
}

\section{The validity of the orbital evolution algorithm}\lb{AppB}
\renewcommand{\theequation}{B.\arabic{equation}}
\setcounter{equation}{0}

\begin{figure}[!t]
	\centering
    \includegraphics[scale =0.80]{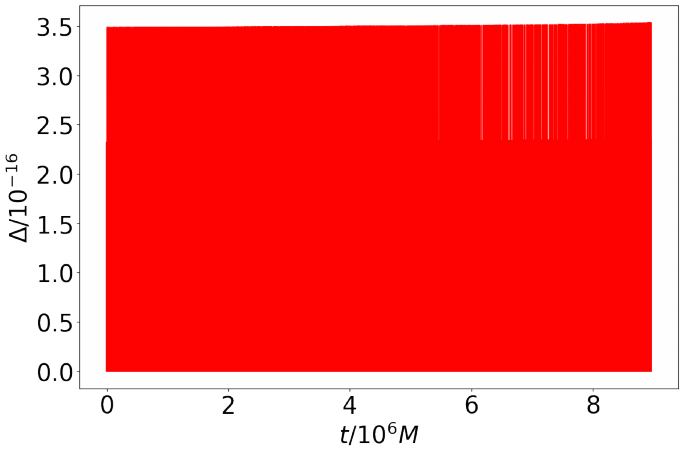}\lb{plot-test-E}
    \caption{The relative numerical
    error between $E(t)$ from gravitational radiation and $\tilde{E}$ from the equations of motion of the object.}
	\label{plot-orbit-E-test}
\end{figure}

To test the validity of the numerical algorithm that we employed to evolve the quasi-circular orbits of a small object, we compare the small object's energy $E(t)$ through the numerical algorithm and energy $\tilde{E}(t)$ from the equations of motion through Eqs. \eqref{Veff-E}, \eqref{Veff}, and \eqref{SCO-condition} over the entire orbital evolution time. For a small object with $m = 10 M_\odot$ inspiraling into a supermassive polymerized black hole with $M = 10^6 M_\odot$ and $\hat{k} = 0.1$, we set the initial position $(r_0, \phi_0) $ of the object as $(10 M, ~ \pi/2)$ and use the orbital evolution algorithm to obtain its complete quasi-circular orbit. We also calculate the evolution of the object's energy and orbital angular momentum as $(E(t), L(t))$ through the orbital evolution algorithm, and obtain another evolution of the object's energy $\tilde{E}(t)$ from the equations of motion. Then we calculate the relative numerical error between $E(t)$ and $\tilde{E}(t)$, which is defined as
\bqn\lb{E-err}
\Delta= \frac{|E - \tilde{E}|}{E}.
\eqn 
The results are shown in Fig.~\ref{plot-orbit-E-test}. It shows that the evolution of the object's energy obtained from Eq.~\eqref{fluex-2} agrees well with that from Eqs.~\eqref{Veff-E}, \eqref{Veff}, and \eqref{SCO-condition}, which indicates that the numerical algorithm we employed to evolve the quasi-circular orbits of a small object is reasonable.

\end{document}